\documentclass[aps,prd,twocolumn,nomerge,superscriptaddress,preprintnumbers,superscriptaddress,nofootinbib,floatfix]{revtex4-1}

\usepackage{amsmath}
\usepackage{graphicx}
\usepackage{siunitx}
\usepackage{comment}
\usepackage{xcolor}

\newcommand{\aj}{\hat{\alpha}_j} 
\newcommand{\hu}{\mathrm{\,km\, Mpc^{-1} s^{-1}}}

\newcommand{\jcap}{Journal of Cosmology and Astroparticle Physics}
\newcommand{\aap}{Astronomy \& Astrophysics}
\newcommand{\apjl}{The Astrophysical Journal letters}
\newcommand{\mnras}{Monthly Notices of the Royal Astronomical Society}
\newcommand{\pasp}{Publications of the Astronomical Society of the Pacific}
\newcommand{\apjs}{The Astrophysical Journal Supplement Series}
\newcommand{\apss}{Astrophysics and Space Science}

\begin{document}

\title{Probing modified gravity theories and cosmology using gravitational-waves and associated electromagnetic counterparts}

\author{S.~Mastrogiovanni, D.A.~Steer and M.~Barsuglia}
\affiliation{AstroParticule et Cosmologie (APC), 10 Rue Alice Domon et L\'eonie Duquet, Universit\'e de Paris, CNRS, AstroParticule et Cosmologie, F-75013 Paris, France}

\begin{abstract}
The direct detection of gravitational waves by the LIGO-Virgo collaboration has opened a new window with which to measure cosmological parameters such as the Hubble constant $H_0$, and also probe general relativity on large scales. In this paper we present a new phenomenological approach, together with its inferencial implementation, for measuring deviations from general relativity (GR) on cosmological scales concurrently with a determination of $H_0$. 
We consider gravitational waves (GWs) propagating in an expanding homogeneous and isotropic background, but with a modified friction term and dispersion relation relative to that of GR.  
We find that a single binary neutron star GW detection will poorly constrain the GW friction term. However, a joint analysis including the GW phase and GW-GRB detection delay could improve constraints on some GW dispersion relations provided the delay is measured with millisecond precision.
We also show that, for massive gravity, by combining 100 binary neutron stars detections with observed electromagnetic counterparts and hosting galaxy identification, we will be able to constrain the Hubble constant, the GW damping term and the GW dispersion relation with 2\%, 15\% and 2 \% accuracy, respectively. We emphasise that these three parameters should be measured together in order avoid biases.
Finally we apply the method to GW170817, and demonstrate that 
for all the GW dispersions relations we consider, including massive gravity, the GW must be emitted $\sim$ 1.74s before the Gamma-ray burst (GRB). Furthermore, at the GW merger peak frequency, we show that the fractional difference between the GW group velocity and $c$ is $\lesssim 10^{-17}$.
\end{abstract} 

\maketitle


\section{Introduction}
A fundamental building block of the Standard Cosmological model ($\Lambda$CDM) is Einstein's theory of general relativity (GR).
When supplemented with the assumption that on large scales the Universe is homogeneous and isotropic (the Cosmological principle), together with the introduction of dark energy --- in the form of a cosmological constant $\Lambda$ --- and dark matter, it is an excellent description of our observable universe, including its accelerated expansion today \cite{2016PDU....12...56B}.  
However, many theoretical questions remain open, most fundamental of which is perhaps the nature of dark energy and dark matter \cite{2011PhRvD..83b3005Z}. 
On the observational side, it is well known that the measurements of the Hubble constant through the Cosmic Microwave Background (CMB) \cite{2016A&A...594A..13P} and in the local Universe with Standard Candles \cite{2019ApJ...886L..27R,2019NatRP...2...10R} show a significant statistical discrepancy \cite{2017NatAs...1E.169F}.
One of the possible solutions to these open problems is to consider that GR is modified on cosmological scales.
There exists numerous models of modified gravity models which break different GR assumptions \cite{2018FrASS...5...44E}, and in general in these models both scalar and tensor perturbations evolve differently from those of GR \cite{2013PhRvD..87l3532A,2019PhRvD..99h3504L,2019PhRvD.100l3501G}. 
In this paper we focus on the tensor perturbations, namely Gravitational Waves (GWs).

Modifying gravity on cosmological scales generally results in a GW dispersion relation, i.e.~GWs do not propagate at the speed of light; and also to a different GW friction term relative to that of GR.
The parameters encoding these deviations are often denoted by $\alpha_M$ (for the friction term) and $\alpha_T$ (for the dispersion relation) \cite{2018JCAP...03..021L}. 
For instance $\alpha_M$ might arise from a running Planck mass, while a non-zero $\alpha_T$ can occur in the case of a massive graviton \cite{2018FrASS...5...44E}.
GWs offers a unique opportunity to probe these parameters on cosmological scales. In fact, from a GW detection is possible to infer directly the luminosity distance of the source \cite{1986Natur.323..310S,2009LRR....12....2S,2005ApJ...629...15H,2010ApJ...725..496N,2008PhRvD..77d3512M} without the use of a cosmological ladder, thus giving the possibility to independently measure $H_0$ with GWs\cite{2012PhRvD..86d3011D,2013arXiv1307.2638N} even if an electromagnetic counterpart is not observed \cite{2019arXiv190806050G}.
Moreover, the GW phase can be  studied to probe for the presence of possible GW dispersion relations \cite{1998PhRvD..57.2061W,2012PhRvD..85b4041M,2016PhRvD..94h4002Y, 2016PhRvL.116v1101A, 2017PhRvD..96j4027S, 2019PhRvD.100j4036A,2019PhRvL.123a1102A}.  Finally, GWs can be also detected together with a gamma-ray burst (GRB), and any delay with respect to the GW can be used for probing $\alpha_T$ \cite{2017ApJ...848L..13A}.

GWs have been used to measure $H_0$ using the GW170817 hosting galaxy identification \cite{2017Natur.551...85A} whereas for binary black holes, galaxy surveys have been employed \cite{2019ApJ...871L..13F,2019arXiv190806060T}. This type of study is indeed very promising for measuring independently  $H_0$, in fact with hundreds of GWs detection we will be able to measure $H_0$ with 2\% accuracy  \cite{2018Natur.562..545C}.
Recently, GWs have been used also to constrain $\alpha_M$ and $\alpha_T$ (or their equivalent quantities) independently of each other,  often without considering the cross-correlations with the Hubble constant $H_0$ (in the case of $\alpha_T$) and never (to the best of our knowledge) considering all of them at the same time.
In \cite{2018JCAP...07..048P,2019arXiv190703150F,2018JCAP...03..021L,2019PhRvD..99h3504L} for instance, the authors present a methodology to measure jointly $\alpha_M$ and $H_0$ from the GW luminosity distance assuming that the redshift of the source is perfectly known from the identification of the host galaxy, while in \cite{2019PhRvD.100d4041D,2020JCAP...03..015B} the authors make use of the luminosity distance of high redsfhit events to measure $\alpha_T$.
In \cite{2018JCAP...07..048P,2019PhRvD..99h3504L,2019PhRvL.123a1102A} a more complete statistical analysis is presented using the binary neutron star merger (BNS) GW170817 \cite{2017PhRvL.119p1101A} for cosntraining the number of spacetime dimensions and a running Planck mass.
Other works such as \cite{2012PhRvD..85b4041M,1998PhRvD..57.2061W} focus on $\alpha_T$ and the GW dispersion relation \cite{2019PhRvL.123a1102A}  by fixing the cosmological parameters, and in particular $H_0$, to the ones measured from CMB \cite{2016A&A...594A..13P}. Others focus GW dispersion relations given by particular gravity models such as Horava gravity \cite{2020arXiv200514705F}, scalar-tensor theories  \cite{2017PhRvL.119y1302C,2018EPJC...78..738G,2020arXiv200606652B} and Horndeski \cite{2018JCAP...12..030K,2019Univ....5..138P,2019IJMPD..2842005K,2020arXiv200605512O} or $f(T/G)$ gravity  \cite{2018PhRvD..97j3513C,2020Ap&SS.365...26K}.
More recently there have been also some modelling effort for considering the effect of $\alpha_T$ and $\alpha_M$ together on the GW waveform \cite{2018PhRvD..97j4037N,2018PhRvD..97j4038A,2019PhRvD..99j4038N,2018PhRvD..98b3504B}, however the cross-correlation with Standard Cosmology are still not considered, the gamma-ray burst emission not taken into account and to the best of our knowledge without a complete inferencial model.

In this paper we present a phenomenological method able to combine GW data, together with its associated GRB and hosting galaxy data to recover a joint estimation of the Hubble constant $H_0$ and the $\alpha_M$ and $\alpha_T$ parameters. We also apply our method to GW170817 and extract joint constraints on these parameters as well as the GW-GRB emission delay.

The paper is organized as follow. 
In Sec.~\ref{sec:2} we discuss the propagation equation for a GW traveling in modified gravity theories. We then link the GR deviation parameters $\alpha_{M/T}$ to three measurable quantities that we can infer from a GW event and for which we are usually provided with posterior samples: the luminosity distance, the GW phase evolution and the GW-GRB delay.
In Sec.~\ref{sec:3} we discuss the level of accuracy needed on these 3 observables to constraint the parameters $H_0, \alpha_M,\alpha_T$ using BNSs mergers observed with Advanced LIGO and Virgo.
In Sec.~\ref{sec:4} we introduce a Bayesian inferencial method which is able to provide a joint estimation of $H_0, \alpha_M,\alpha_T$ starting from the posteriors provided for the 3 observables and taking into account redshift uncertainties.
In Sec.~\ref{sec:5} we validate our method with simulated posterior samples, and we discuss the accuracy that we can reach on $\alpha_M, \alpha_T, H_0$  on combining $\mathcal{O}(100)$ BNSs detections. We show that in order to not obtain a biased measure, these parameters should always be considered together.
In Sec.\ref{sec:GW170817} we apply our method to the case of GW170817 and explain its implications.
Finally in Sec.~\ref{sec:6} we draw our conclusions and the prospects for this work.

\begin{figure} 
    \centering
    \hspace{-2cm}
    \includegraphics[scale=.35]{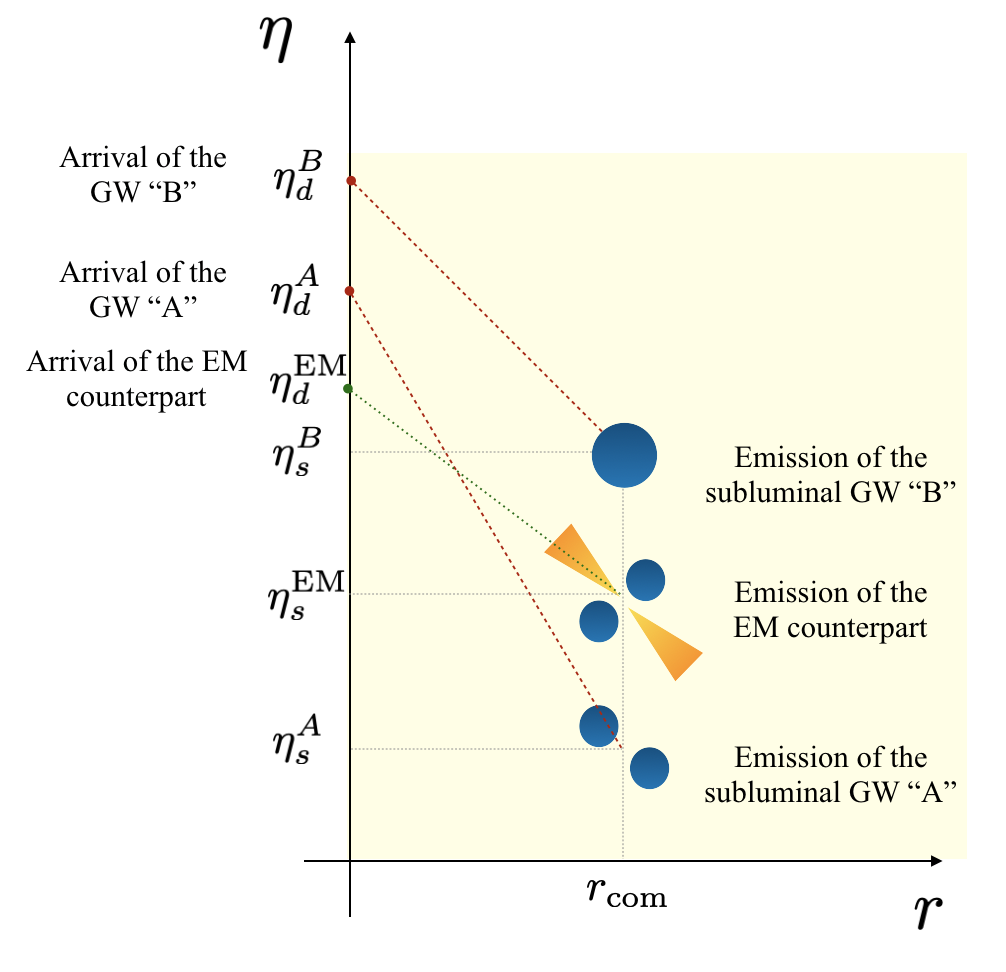}
    \caption{Schematic figure showing a binary neutron star merger at comoving distance $r_{\rm com}$ emitting a GW and a GRB. The blue dots represent the two binary neutron stars. In modified gravity, GWs may propagate with a frequency dependent speed, and arrive with a relative time delay with respect to the electromagnetic counterpart. In this example, photons follow a null-like geodesic identified by and angle of $45$ deg on the plot. Subluminal GWs follow a time-like geodesic identified by an angle lower than $45$ deg (angle defined counterclockwise between $r-\eta$ axes).}
    \label{fig:sketch}
\end{figure}

\section{Propagation in modified gravity \label{sec:2}}
We consider a flat Friedman-Lemaitre-Robertson-Walker (FLRW) background space-time with line element
\begin{equation}
    ds^2 = -c^2 dt^2 + a^2(t) d\vec{x}^2
\end{equation}
where $a(t)$ is the scale factor and conformal time $\eta$ is defined by
\begin{equation}
    \eta = \int \frac{dt}{a(t)}.
\end{equation}
Though our focus is on modified gravity theories, we assume that the background evolution of the scale factor $a(t)$ is as in standard $\Lambda$CDM cosmology, so that only the dynamics of perturbations are modified relative to those of general relativity (a standard approximation on the literature.)  Here we focus on tensor perturbations, namely gravitational waves. See for instance \cite{2013PhRvD..87j4015S} for a discussion of scalar perturbations.

Consider a source at a fixed comoving position $r_{\rm com} = |\vec{x}|$, which emits both a light (GRB) and a GW, see figure \ref{fig:sketch}.  
Light rays travel along null geodesics and hence, assuming the observer being at the origin, $c(\eta_d^{\rm EM} - \eta_s^{\rm EM}) = r_{\rm com}$, where $\eta_d^{\rm EM}$ ($\eta_s^{\rm EM}$) is the conformal time at which EM waves are detected (emitted). Light rays are redshifted in the usual way by the cosmological expansion;
\begin{equation}
f^{\rm EM}_{s}= (1+z) f^{\rm EM}_{d},
\end{equation}
where
\begin{equation}
1+z = \frac{a(\eta^{\rm EM}_d)}{a(\eta^{\rm EM}_s)}.
\label{eq:zdef}
\end{equation}
 
As indicated in Fig.~\ref{fig:sketch}, in modified gravity models,  GWs can travel with (possibly frequency dependent) speed $c_T \neq c$ and
furthermore, as we discuss below, they are generally subjected to a modified friction term relative to that of general relativity (see \cite{2019JCAP...07..024B,2018PhRvD..97j4066B,2018PhRvD..98b3510B,2019JCAP...08..015B,2018PhRvD..98b3504B} for some reviews and examples). 
Indeed here, as in many other papers in the literature, we only consider the effect of the modified gravity on the {\it propagation} on the GW signal \cite{1998PhRvD..57.2061W,2012PhRvD..85b4041M,2019PhRvD..99j4038N,2018PhRvD..97j4038A}. That is, the wave GW signal {\it emitted} by the binary source is assumed to be given by the standard GR expression. 
Hence we also restrict our analysis to the standard 2 degrees of freedom of GR\footnote{In many modified gravity models there are  more than  2 propagating polarisations \cite{2004PhRvD..70b4003J}, which may possibly interact with each other. Here do not consider this case but focus on the effect of a modified propagation speed and friction term}. 

Our starting point is a modified dispersion relation of the form \cite{2012PhRvD..85b4041M} 

\begin{equation}
c^2 g_{\mu \nu} p^{\mu} p^\nu =- {B_\alpha} |cp|^\alpha.
\label{eq:dispersion}
\end{equation}
 where, for GWs emitted at $r_{\rm com}$ and propagating radially to the observer, 
\begin{equation}
    p^\mu = (E/c, \hbar k/a^2,0,0)  
\end{equation}
with $k$ the (constant) comoving wave number, and $|p| = (g_{ij} p^i p^j)^{1/2} =\hbar k/a^2$.  Thus the dispersion relation \eqref{eq:dispersion} is
\begin{equation}
{E^2} = {c^2}\frac{\hbar^2 k^2}{a^2} + {B_\alpha} \left(c\frac{ \hbar  |k|}{a}\right)^\alpha,
\label{eq:dispersion2}
\end{equation}
which depends on the physical momentum $p_{\rm ph} = k/a$.  When the coefficients $B_\alpha$ vanish, the dispersion relation Eq.~\eqref{eq:dispersion} reduces to the standard one of a massless particle in general relativity $\omega \equiv E/\hbar = c k/a$. For $B_0 \neq 0$, Eq.~\eqref{eq:dispersion2} is the dispersion relation for the massive graviton $B_0=m_g^2 c^4$ (in [eV]$^2$).  
Different theories give different predictions for the (generally $\eta$-dependent) $B_\alpha$, see \cite{2012PhRvD..85b4041M} for some examples. Here we aim to see what constraints GW observations can put on the $B_\alpha$ without focusing on any particular theory. 

Let us rewrite Eq.~\eqref{eq:dispersion2} as
\begin{equation}
    E^2 \equiv \hbar^2 \omega^2 = c_T^2(\eta,k/a) \frac{\hbar^2 k^2}{a^2}
    \label{eq:dispersion3}
\end{equation}
where 
\begin{equation}
   c^2_T(\eta,k/a) \equiv c^2 \left[1+ {B_\alpha} \left(c\frac{ \hbar  |k|}{a}\right)^{\alpha-2}\right].
   \label{eq:phasevel}
\end{equation}
Motivated by the very tight constraint on the speed of of gravitational waves \cite{2003LRR.....6....5S,2019PhRvL.123a1102A}, we will assume that GWs are ultra-relativistic and that 
\begin{equation}
|B_\alpha|\left(c\frac{ \hbar  |k|}{a}\right)^{\alpha-2}   \ll 1.
\label{eq:nonR}
\end{equation}
Then from Eq.~\eqref{eq:dispersion3} it follows that
\begin{equation}
\omega \simeq c|k|/a,
\label{eq:approxE}
\end{equation}
so that the frequency of the emitted GW $f^{\rm GW}_s$ is related to that of the observed GW $f^{\rm GW}_d$ by the standard redshift relationship, namely
\begin{equation}
a(t_d) f^{\rm GW}_d \simeq a(t_s) f^{\rm GW}_s.
\end{equation}
Hence we can identify the the GW redshift with the usual photon redshift $z$, see Eq.~\eqref{eq:zdef}.
With this approximation 
\begin{equation}
    k \approx \frac{1}{c} \omega(\eta_d)a(\eta_d)=2\pi \frac{f_d}{c}
    \label{eq:approx1}
\end{equation}
since today $a=1$.
This allows us to write the phase velocity in Eq.~\eqref{eq:phasevel} in terms of the detected GW frequency $f_d$; 
\begin{equation}
c^2_T(\eta,f_d/a) = c^2 \left[1+\hat{\alpha}_j \left(\frac{ f_d}{a} \right)^{j}\right],
\label{eq:modg}
\end{equation}
where we have defined 
\begin{equation}
    \hat{\alpha}_j=B_{j+2}(2\pi \hbar)^j
\end{equation}
with $j=\alpha-2$.  Notice that the dimensions of $[\alpha_j]={\rm Hz}^{-j}$.
The radial propagation velocity of the waves is given by 
\begin{equation}
    \frac{dr}{dt} = \frac{p^r}{p^t}= c^2 \frac{k}{a} \frac{1}{a\omega} =\frac{v_g}{a}=\frac{1}{a} \frac{dr}{d\eta}
\end{equation}
where
the group velocity 
\begin{equation}
    v_g \simeq c \left[1-\frac{\hat{\alpha}_j}{2} \left(\frac{f_d}{a}\right)^{j}\right],
    \label{eq:vg}
\end{equation}
and we have used the approximation Eq.~\eqref{eq:nonR}. For example, for massive gravitons $j=-2$, and $c_T>c$, but the group velocity $v_g$ is smaller than $c$.

The dispersion relation in Eq.~\eqref{eq:dispersion3} can be obtained from the wave equation\footnote{This assumes that $a$ and $B_\alpha$ varies on a cosmological time scale, which is much larger than any time-scale associated with the GW. Or in terms frequency (and in natural units), $1/k \ll r_{\rm{com}} \ll H_0^{-1}$.}

\begin{equation}
    \chi''(\eta,k) + k^2c^2_T (\eta,k/a) ^2 \chi(\eta,k)=0
    \label{eq:wavechi}
\end{equation}
where $'=d/d\eta$ and $\chi$ is the radial component of the propagating wave.
The GW perturbation $h$ (we drop the tensor indices for the moment) is related to $\chi$ through (see e.g.~\cite{2018PhRvD..97j4066B})
\begin{equation}
    \chi = \tilde{a} h.
    \label{chidef}
\end{equation}
Here $\tilde{a}$ is an effective scale factor that encodes additional modifications to the GW friction term. We parameterize it as 
\begin{equation}
\frac{\tilde{a}'}{\tilde{a}} \equiv [1+\alpha_M (\eta)] \frac{a'}{a}
\label{eq:tildea}
\end{equation}
where $\alpha_M(\eta)$ is a deviation factor that can parameterize several theories such as scalar-tensor theories with a running Planck mass or theories with extra-dimensions. On subhorizon scales (that is, on scales smaller than $\tilde{a}''/\tilde{a}$ \cite{2019JCAP...08..015B}), Eq.~\eqref{eq:wavechi} can be obtained from 
\begin{equation}
    h''+2[1+\alpha_M (\eta)] \frac{a'}{a}h' + k^2c_T^2(\eta,k/a) h=0,
    \label{eq:FRWa}
\end{equation}
which is the wave equation of a GW propagating with a modified dispersion relation in the FRLW universe. We can solve it using the WKB approximation following \cite{2018PhRvD..97j4037N,2019PhRvD..99j4038N}, and obtain \cite{2018PhRvD..98b3504B}
\begin{equation}
h(\eta,k) = h_{{\rm GR}} (\eta_s,k) C(\eta,\eta_s,k).
\end{equation}
where $h_{{\rm GR}} (\eta_s,k)$ is the solution in GR at the source at comoving distance $r_{\rm com}$ , and $C$ can be interpreted as the transfer function from the source to the detector for each GW mode $k$ .
In terms of conformal time and detected GW frequency $f_d$ (recall from Eq.~\eqref{eq:approx1} that $k\simeq 2\pi f_d/c$) it is given by
\begin{eqnarray}
 C(\eta,\eta_s,k) &=& \left[\frac{c_T(\eta_s,f_d/a(\eta_s))}{c_T(\eta,f_d/a(\eta))}\right]^{1/2}  \frac{\tilde{a}(\eta_s)}{\tilde{a}(\eta)} \times \nonumber \\ && {\rm exp}[2\pi i (f_d/c)\int_{\eta_s}^\eta c_T(\eta',f_d/a) d\eta']
 \nonumber \\  &\equiv &|C(\eta,\eta_s,f_d)|e^{i \Psi(\eta,\eta_s,f_d)}.
\label{eq:transfer}
\end{eqnarray}
The modulus of $C$ will contribute to the GW amplitude, that is to a  modification of the luminosity distance. Its phase $\Psi(\eta,\eta_s,f_d)$ leads to time delays and phase shifts, as we now discuss. 

\subsection{Observables}

\subsubsection{Luminosity distance}
The first estimator that we define arises from the modulus of the transfer function.
In GR, the amplitude of the GW scales as the comoving distance of the source. From Eq.~(\ref{eq:transfer}), in modified gravity, the GW amplitude at the detector is is now given by
\begin{equation}
d^{\rm GW}(\eta_d,f_d) = r_{\rm com} \frac{\tilde{a}(\eta_d)}{\tilde{a}(\eta_s)} \left[\frac{c_T(\eta_d,f_d/a(\eta_d))}{c_T(\eta_s,f_d/a(\eta_s))} \right]^{1/2}.  
\label{eq:GWlum1}
\end{equation}
Since the results on GW dispersion relations are very tight $|c-c_T|<10^{-15}$ \cite{2019PhRvL.123a1102A, 2019PhRvD.100j4036A}, and measured errors on $d^{\rm GW}$ are typically of at least a few percent, usually the effect of $c_T$ on the distance is negligible. This is also consistent with the assumption in Eq.~\eqref{eq:approxE}.
The term $\tilde{a}
$ encodes the deviations in the GW friction and from Eq.~\eqref{eq:tildea}, using redshift instead of conformal time, we obtain
\begin{equation}
    \tilde{a}(z)= a(z) \mathrm{exp} \left[-\int_{0}^{z} \frac{\alpha_M (z)}{1+z}dz\right],
\end{equation}
where we  have assumed that $a(0)=\tilde{a}(0)=1$.
In terms of the standard luminosity distance  $d_{\rm EM}(z) = r_{\rm com} /a(\eta_s)=r_{\rm com} (1+z)$, 
we find that the GW luminosity distance in modified gravity is given by
\begin{equation}
d^{\rm GW}(z) = d_{\rm EM} (z) \mathrm{exp} \left[\int_{0}^{z} \frac{\alpha_M (z)}{1+z}dz\right]. 
\label{eq:GW_lum}
\end{equation}
This equation is consistent with previous works \cite{2019JCAP...07..024B,2018PhRvD..97j4066B,2018PhRvD..98b3510B,2019JCAP...08..015B}, which have shown the potential of the modified luminosity distance to be a good marker for testing possible deviations from GR on cosmological scales.

\begin{figure}[h!]
    \centering
    \includegraphics[scale=0.5]{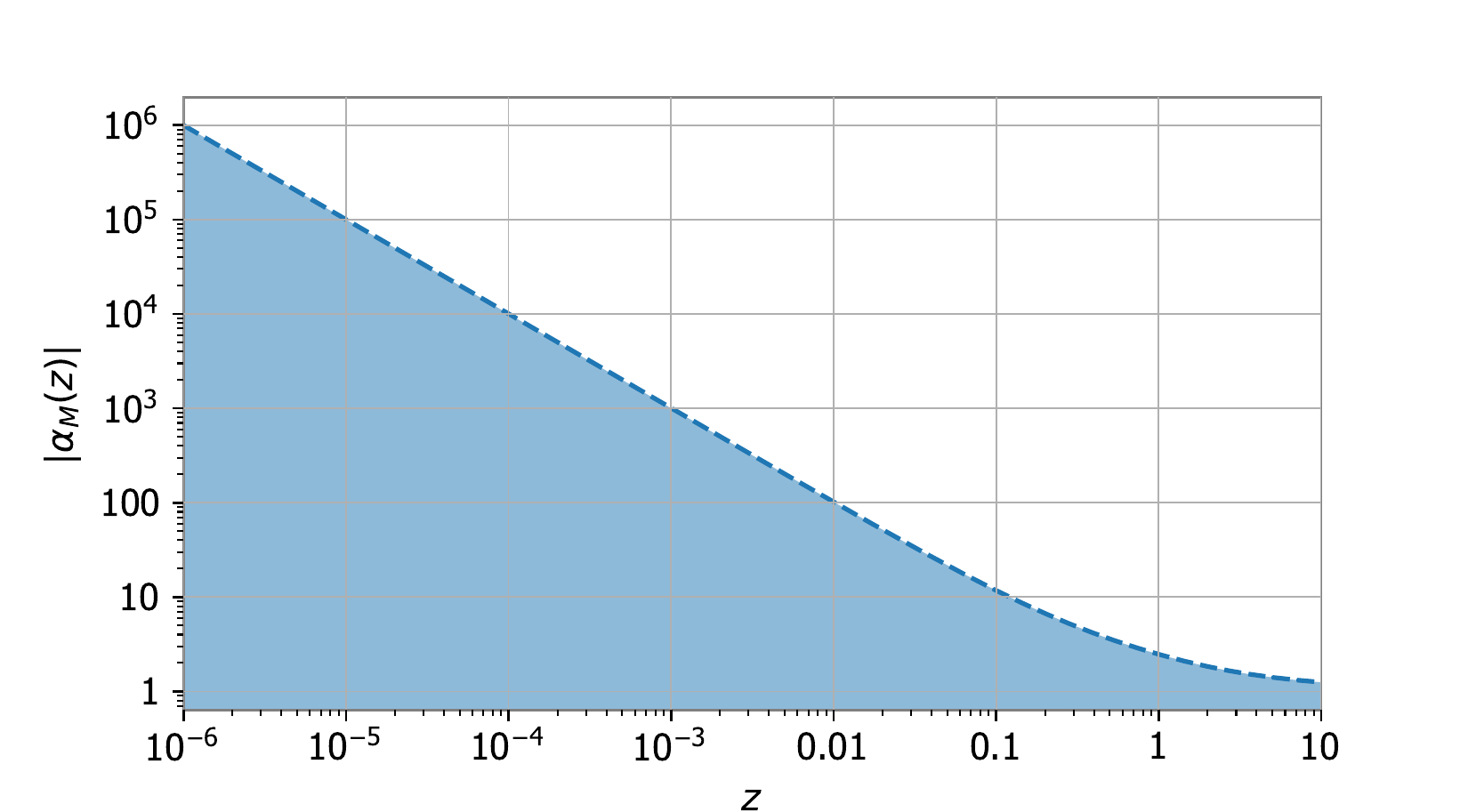}
    \caption{The shaded area of on the plot shows the allowed value for the parameter $\alpha_M$ with respect to the redshift. Any functional form of $\alpha_M$ in the shaded area, will result in a monothonically increasing GW luminosity distance.}
    \label{fig:alpha_condition}
\end{figure}

We now deviate from these references and use Eq.~\eqref{eq:GW_lum} to bound the parameter $\alpha_M (z)$ such that the GW luminosity distance is a monotonically increasing function of the redshift.  
This condition is physically motivated, since if it were not satisfied one would detect an infinite number of GWs sources at higher redshifts. In order to avoid this unphysical case, $\alpha_M$ must satisfy
\begin{equation}
    \alpha_M (z) \geq  -\frac{(1+z)}{E(z)} \left[ \int_0^z \frac{dz'}{E(z)} \right]^{-1} -1,
    \label{eq:al_con}
\end{equation}
where 
\begin{equation}
    E(z) = \sqrt{\Omega_M (1+z)^3 + \Omega_\Lambda}.
\end{equation}
Since the right hand side of Eq.~\eqref{eq:al_con} is negative it follows that any positive values of $\alpha_M$ (corresponding to a further GW), will satisfy this condition. Of course this is not valid for negative values of $\alpha_M$ (GW might appear closer.)
Fig.~\ref{fig:alpha_condition} shows the allowed values for GW friction parameter $\alpha_M$ computed with Planck values of $\Omega_M=0.308$ \cite{2016A&A...594A..13P} and $\Omega_\Lambda=1-\Omega_M$.
Since at lower redshifts the $\alpha_M$ contribution to the GW luminosity distance is small, this term is allowed to take very large values. However at higher redshifts, $\alpha_M$ must be constrained to smaller values in order to satisfy the condition in Eq.~\eqref{eq:al_con}.

\subsubsection{Time delay}

We now compute the time delay at the {\it detector} between two monochromatic GWs which were emitted at different times from the source at fixed comoving distance $r_{\rm com}$, see Fig.~\ref{fig:sketch}.  Consider a GW emitted at $\eta_s^A$ and received at $\eta_d^A$, with detected frequency $f_{d,A}$. From Eq.~\eqref{eq:vg} it follows that
\begin{equation}
    r_{\rm com} = \int_{\eta_s^A}^{\eta_d^A} c\left[1-\frac{1}{2}\hat{\alpha}_j \frac{f_{d,A}^j}{a^j} \right]d \eta.
    \label{thisone}
\end{equation}
Similarly for second GW labelled by $B$ we have
\begin{equation}
    r_{\rm com} = \int_{\eta_s^B}^{\eta_d^B} c\left[1-\frac{1}{2}\hat{\alpha}_j \frac{f_{d,B}^j}{a^j} \right]d \eta.
\end{equation}
Hence the conformal time delay at the detector between the two GWs A and B is given by\footnote{We assume $\int_{\eta_s^A}^{\eta_d^A} \hat{\alpha}_j (\eta)/a^j d \eta \approx \int_{\eta_s^B}^{\eta_d^B} \hat{\alpha}_j (\eta)/a^j d \eta$}
\begin{equation}
     \Delta \eta_d^{\rm AB} = \Delta \eta_s^{\rm AB} + \frac{f_{d,A}^j-f_{d,B}^j}{2} \int_{\eta_s^A}^{\eta_d^A} \hat{\alpha}_j (\eta) \left(\frac{1}{a}\right)^{j} d \eta.
\end{equation}
To a good approximation the two GWs are emitted 
and detected on timescales which are smaller than the cosmological timescale, and thus 
\begin{equation}
     \Delta t_d^{\rm AB} = (1+z_s) \Delta t_s^{\rm AB} + \frac{f_{d,A}^j-f_{d,B}^j}{2} \mathcal{T}_{j}
     \label{eq:GWGWtimedelay}
\end{equation}
where
\begin{equation}
\mathcal{T}_{j}=\int_0^{z_s} dz' \hat{\alpha}_j (z) \frac{(1+z')^j}{H_0 E(z')}
\label{eq:tauk}
\end{equation}
and $z_s$ is the fixed source redshift.

In GR $\hat{\alpha}_j=0= \mathcal{T}_{j}$ and Eq.~\eqref{eq:GWGWtimedelay} reduces to the standard time dilation due to cosmological expansion.
Although time delays between different GW modes are not usually measured (see below for a discussion of the measured GW phase), we can exploit Eq.~\eqref{eq:GWGWtimedelay} to obtain another observable quantity, namely the GW-photon time delay.
To do so we set the non-GR delay contribution of $B$ to zero, so that the time delay between the GW its associated electromagnetic counterpart (a GRB) is
\begin{eqnarray}
   \Delta t_d^{\rm GW-EM} &=& (1+z_s) \Delta t_s^{\rm GW-EM}
   + \frac{f_{R,d}^j}{2} \mathcal{T}_{j},
     \label{eq:GWGRBtimedelay}
\end{eqnarray}
where $f_{R,d}$ is understood to be the GW reference frequency used to compute the time delay (For instance for GW170817, $f_{R,d}$ was the merger frequency \cite{2017ApJ...848L..13A}) and $\Delta t_s^{\rm GW-EM}$ is the prompt time delay of the GW and its EM counterpart at the source.

\subsubsection{GW phase shift}

Usually at the detector we do not measure the delay between different GW frequencies but rather the phase of the GW $\psi(f_d)$. This is well approximated by \cite{2008APS..APRJ10009H,2016PhRvD..93l4004S,1999PhRvD..59l4016D}
\begin{equation}
    \psi (f_d)=  2 \pi \int_{f_{R,d}}^{f_d} (t_d-t_{R,d})df_d' +2 \pi f_d t_{R,d} - \phi_{R,d}-\pi/4,
\end{equation}
where $t_{R,d}$ is the reference time at the detector at which the GW had frequency $f_{R,d}$.
Substituting $(t-t_{R,d})$ with the time-delay term computed in Eq.~\eqref{eq:GWGWtimedelay}, and integrating gives
\begin{eqnarray}
     \psi(f_d)&=&\psi_{\rm GR}(f_d)+\pi \mathcal{T}_j\frac{f_d^{j+1}}{j+1},\quad \text{ when} \, j\neq -1
     \label{eq:aGW_dep}
     \\
     \psi(f_d)&=&\psi_{\rm GR}(f_d)+\pi \mathcal{T}_j \ln{f_d}, \quad \text{ when} \, j=-1,    \label{eq:aGW_dep2}
\end{eqnarray}
where
 \begin{equation}
  \psi_{\rm GR}(f_d) = 2\pi f_d t'_{R,d} - \phi'_{R,d} + 2\pi  \int^{f_d}_{f_{R,d}} (1+z_s) (t_s-t_{R,s}) df_d'.
  \nonumber
  \end{equation}
(Here the primes take into account the redefinition of the reference time and phase after the integration on the non-GR delay term \cite{2012PhRvD..85b4041M}.) 
Note that Eq.~\eqref{eq:aGW_dep} predicts a constant phase shift for the non-GR model with $j=0$. This is normal since for $j=0$ the GW group velocity in Eq.~\eqref{eq:vg} is frequency independent.

Usually $\psi_{\rm GR}(f_d)$ is measured in terms of Post-Newtonian (PN) coefficients $\beta^{\rm PN}_n, \beta^{\rm PN}_{n, {\rm ln}}$
\begin{eqnarray}
\psi_{\rm GR}(f_d)&=&\sum_n \psi_{n,{\rm GR}}(f_d)
\nonumber
\\
&=&\sum_n [\beta^{\rm PN}_n +\beta^{\rm PN}_{n, {\rm ln}} \ln f_d] f_{d}^{(n-5)/3}.
\label{eq:PN_exp}
\end{eqnarray}
These may be written in the form \cite{2009PhRvD..80h4043B} 
\begin{equation}
    \beta^{\rm PN}_n = \frac{3}{128 \nu} (\pi M)^{\frac{n-5}{3}} g_{n}(\nu,S_1,S_2),
    \label{eq:PNexp}
\end{equation}
where $M=m_1+m_2$ is total mass of the binary in seconds, $\nu=(m_1 m_2)/M^2$ is the symmetric mass ratio, $S_1,S_2$ the reduced spins and $g_n (\nu,S_1,S_2)$ are numerical functions provided in \cite{2009PhRvD..80h4043B}.
By comparing the frequency dependency of Eqs.~(\ref{eq:aGW_dep}) and (\ref{eq:PN_exp}), we therefore see that a modified dispersion relation with power $j$ will appear as a GR PN parameter of order 
\begin{equation}
    n=3j+8.
\end{equation}
GW waveforms are known up to the PN orders n=0,...,7 thus meaning we can probe dispersion relations with eight different powers $j$ given by
$j=-8/3,-7/3,\ldots,-1/3$
Amongst these cases is massive gravity for which $j=-2$.

With current GW posteriors, there exists two methodologies for probing GW dispersion relations. The first one provides posteriors for GW dispersion relations which are not directly linked to the PN parameters \cite{2012PhRvD..85b4041M}. This method  fixes  $H_0$ to Planck's cosmology \cite{2016A&A...594A..13P} and use the entire GW signal.
The other one, is more agnostic and provides posteriors on the fractional phase deviations in the GW phase that would correspond to a given PN order \cite{2012PhRvD..85h2003L,2014PhRvD..89h2001A} from the inspiral part of the merger. 
In this paper, we start from the agnostic posterior samples on the PN parameters in \cite{2012PhRvD..85h2003L,2014PhRvD..89h2001A}, since we would like to analyse GW dispersion relation connection to the PN coefficients $\beta^{\rm PN}$.
\begin{eqnarray}
     \frac{\psi_{3j+8}(f_d)-\psi_{3j+8,{\rm GR}}(f_d)}{\psi_{3j+8,{\rm GR}}(f_d)}&=&\pi \frac{\mathcal{T}_j}{\beta^{\rm PN}_{3j+8}(j+1)},\, 
     \nonumber
     \\
    && \qquad  \text{ when} \, j\neq -1
     \label{eq:GW_dep}\\
     \frac{\psi_5(f_d)-\psi_{5,{\rm GR}}(f_d)}{\psi_{5,{\rm GR}}(f_d)}&=&\pi \frac{\mathcal{T}_{-1}}{\beta^{\rm PN}_{5, {\rm ln}} },
     \nonumber
     \\
    && \qquad  \text{ when} \, j=-1.     \label{eq:GW_dep2}
\end{eqnarray}

\section{Impact of GR deviations on the three observables \label{sec:3}}

In this section we study the accuracy to which measurements of our three observables --- luminosity distance, time delay and phase shift --- are required in order to make constraints on deviations from GR. 
These deviations are encoded in the $\alpha_M$ and $\hat{\alpha}_j$ parameters; these two parameters, together with the Hubble constant $H_0$ (which is a third parameter) will affect the three GW related observables. From now on, in this paper, we will use the 0th order Taylor's expansion of $\alpha_M(z) \approx \alpha_M(0) ,\hat{\alpha}_j(z) \approx \hat{\alpha}_j(0)$ since we are looking at events at very low redshift. 

As a first step, in this section we discuss the possibility of measuring $\alpha_M$ and  $\hat{\alpha}_j$ from the three observables independently when {\it fixing} $H_0$. 
This will help us anticipate the results on the inference for $\alpha_M, \hat{\alpha}_j$ and $H_0$ which we will show for our statistical method (introduced in Sec.~\ref{sec:4}) applied to software simulated signals in Sec.~\ref{sec:5}. 
In the following we will assume a GW detection horizon (this means that every GW emitted within this range is detected with probability 1) of 333 Mpc and the values of $\alpha_M$ are limited through the condition given in Eq.~\eqref{eq:al_con}. 
The GW horizon we consider roughly corresponds to the BNS range for a 3 detector network with same sensitivity during the observing run 4 (O4) scenario \cite{2018LRR....21....3A}.

\subsection{Measuring $\alpha_M$ from the GW luminosity distance}

As seen from Eq.~\eqref{eq:GW_lum}, the parameter
$\alpha_M$ appears in the GW luminosity distance in Eq.~\eqref{eq:GW_lum}. 
We define the fractional change introduced by the parameter $\alpha_M$ on the GW luminosity distance as
\begin{equation}
    \epsilon_d\equiv\frac{d_{\rm GW}-d_{\rm EM}}{d_{\rm EM}}={\rm{exp}}\left[\int_0^{z_s} \frac{\alpha_M(z) dz'}{1+z'}\right]-1
    \label{eq:eps_d}.
\end{equation}
In principle, in order to measure the contribution of $\alpha_M$, one would need an accuracy on the GW luminosity distance measurement of the same order of the deviation introduced by the GR deviation parameter (assuming we perfectly know $H_0$ and the source redshift).

\begin{figure}[h!]
    \centering
    \includegraphics[scale=0.5]{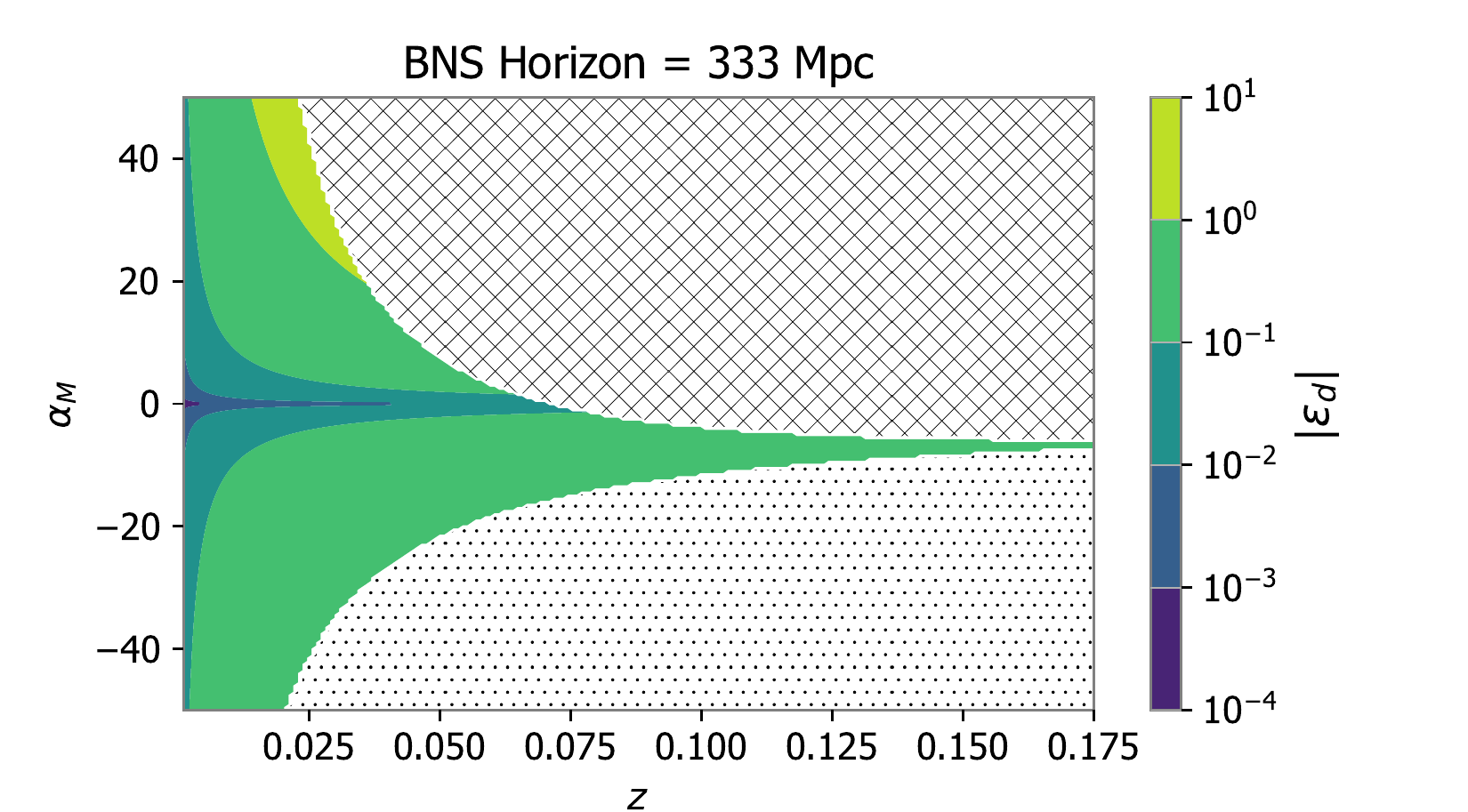}
    \caption{Absolute value of the fractional change (colorbar) on the GW luminosity distance introduced by $\alpha_M$ at different redshifts for a BNS detection during O4 observing scenario. We assume a value for the Hubble constant of $H_0=70 {{\rm\, km\, Mpc}^{-1}s^{-1}}$. The hatched region corresponds to BNS that cannot be detected since they are above the GW horizon. The dotted region is excluded from Eq.~\eqref{eq:al_con}}
    \label{fig:lum_GW_change_horizon_BNS_333Mpc}. 
\end{figure}

In Fig.~\ref{fig:lum_GW_change_horizon_BNS_333Mpc}, we show the absolute value of the GW luminosity distance fractional change, $\epsilon_d$, for different source redshifts and varying values of $\alpha_M$.
In this case study, we fix the value of the Hubble constant to $H_0=70 {{\rm\, km\, Mpc}^{-1}s^{-1}}$.

We see that a precision on the GW luminosity distance between 10\% and 1 \% is needed in order to constrain $\alpha_M$ between -5 and 5 for sources detected between redshifts $0.01$ and $0.06$.
Unfortunately, this accuracy is not achievable with current GW detections; indeed, most of the error budget for the luminosity distance inferred from GW data comes from the well-known distance-binary inclination degeneracy\cite{2018PhRvL.121b1303V,2019PhRvX...9c1028C},
and as a result, typical values of the luminosity distance uncertainty are of the the order of 20\%-40\% \cite{2019PhRvD.100h3514C} depending on the detected signal-to-noise ratio.
This means at the best  we expect to constrain the $\alpha_M$ parameter between values of order  of -20 and 20 for BNSs between redshift 0.02 and  0.04. 
For BNSs detected at redshifts around 0.01 (like GW170817), the constraint will be of the order of -40 and 40. 
This prediction is consistent with the value of $\alpha_M$ found in \cite{2019PhRvD..99h3504L} from GW170817. As we will also see later in Sec.~\ref{sec:5}, a better accuracy will be reached combining the results from many GW detections.

\begin{figure*}[h!]
    \centering
    \includegraphics[scale=0.5]{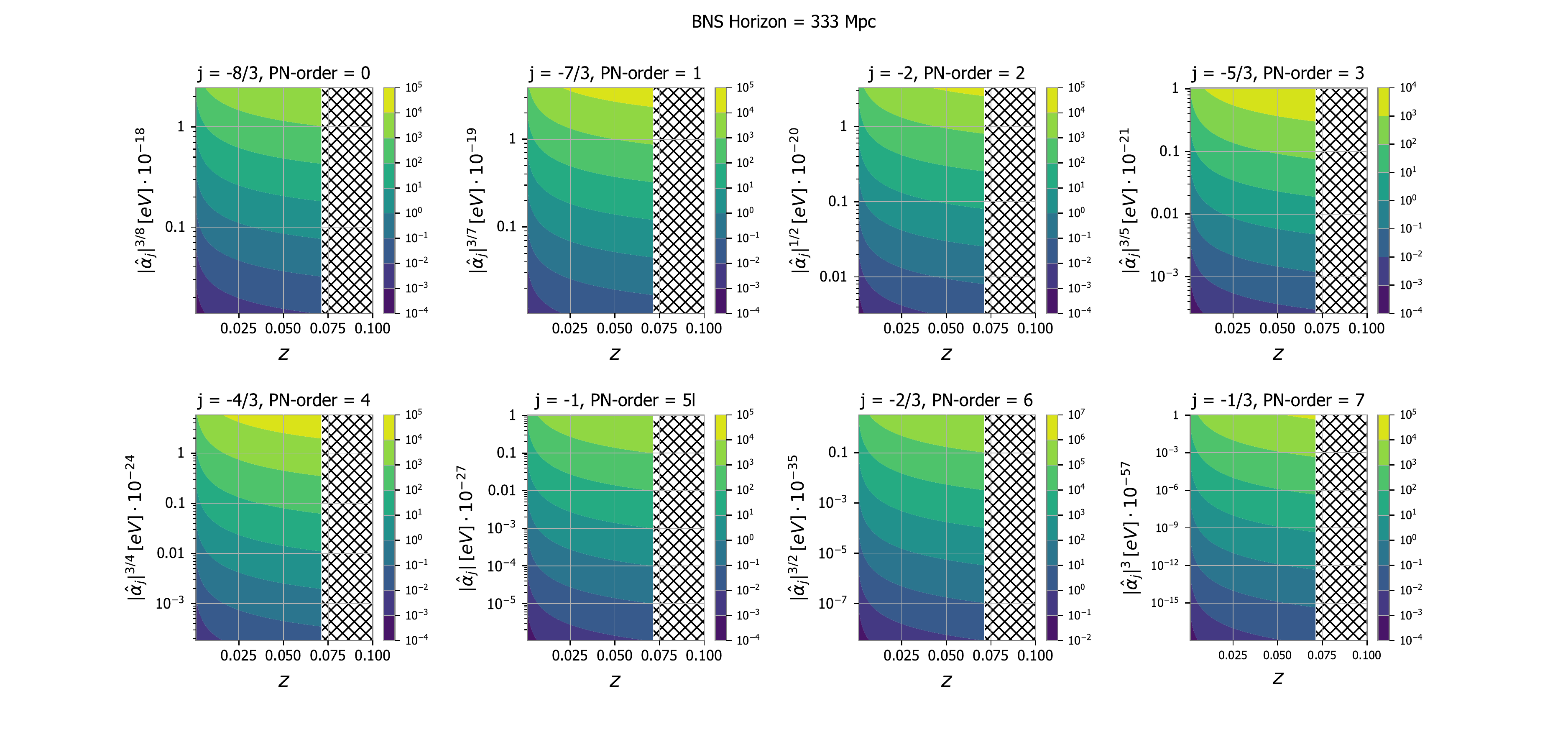}
    \caption{Fractional deviation in the PN coefficients introduced by the GW dispersion parameter $\hat{\alpha}_j$ with respect to the source redshift for a BNS O4 observing scenario. The Hubble constant has been fixed to $H_0=70 {{\rm\, km\, Mpc}^{-1}s^{-1}}$. The hatched area corresponds to sources above the O4 GW horizon for detectability, and in each case the range of the $y$-axis satisfies (\ref{eq:cond_t}). }
    \label{fig:BNS_dphi_difference}
\end{figure*}

\begin{figure*}[h!]
    \centering
    \includegraphics[scale=0.5]{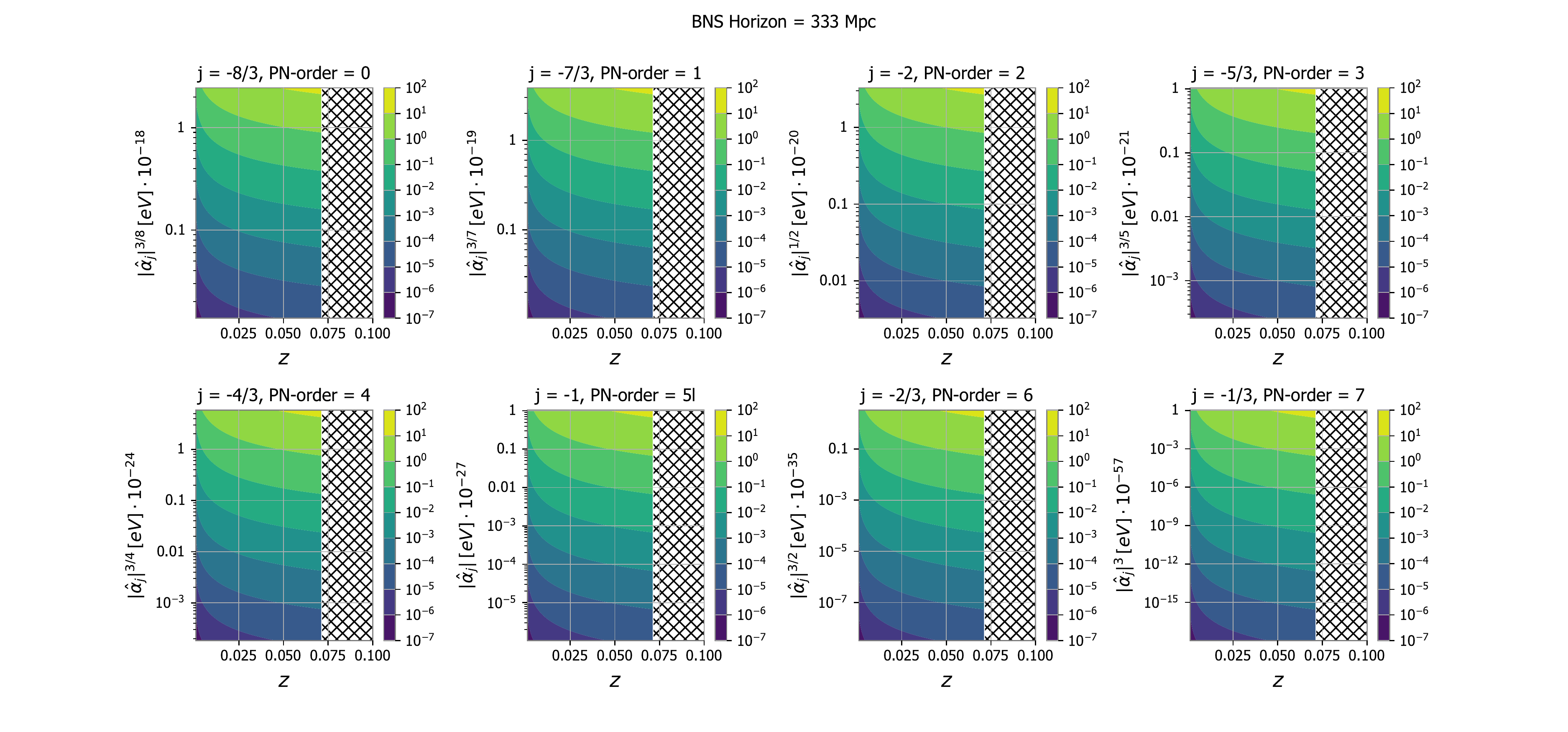}
    \caption{Time delay in seconds (colorbar) introduced by the GW dispersion parameter $\hat{\alpha}_j$ with respect to the source redshift for a BNS O4 observing scenario. The Hubble constant has been fixed to $H_0=70{{\rm\, km\, Mpc}^{-1}s^{-1}}$. The hatched area corresponds to sources above the O4 GW horizon for detectability, and in each case the range of the $y$-axis satisfies (\ref{eq:cond_t}).}
    \label{fig:BNS_dt_difference}
\end{figure*}

\begin{figure*}[htp!]
    \centering
    \includegraphics[scale=0.5]{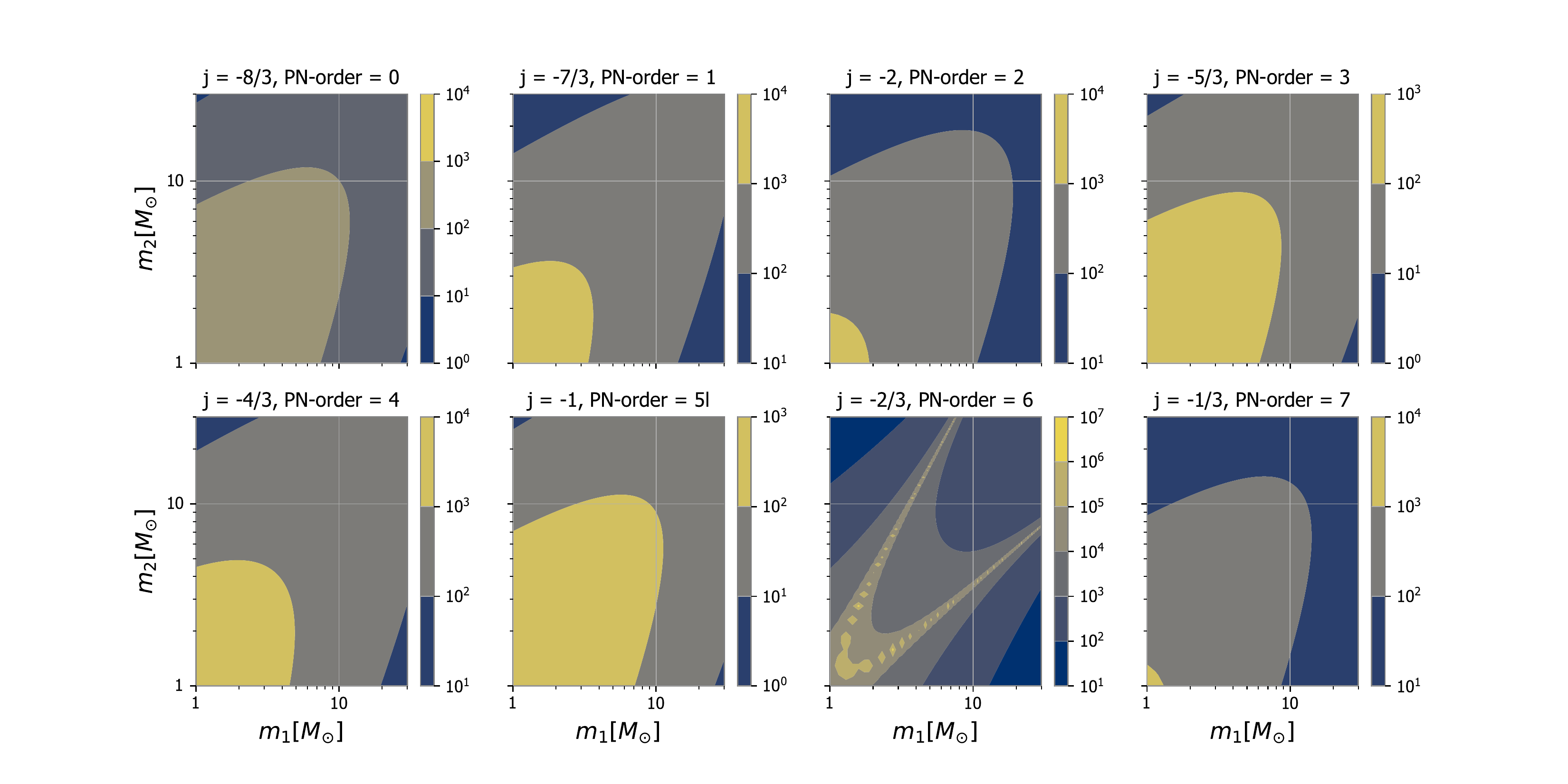}
    \caption{Ratio of the fractional changes introduced on the GW phase and GW time delay with a  GW dispersion relation.}
    \label{fig:phi_accuracy}
\end{figure*}

\subsection{Measuring $\hat{\alpha}_j$ from the GW phase and the GRB delay}

We again consider the O4 BNS observing scenario and a value of $H_0=70 {{\rm\, km\, Mpc}^{-1}s^{-1}}$. Furthermore, in this subsection, in order to gain intuition on the parameter $\hat{\alpha}_j$, we fix $\alpha_M = 0$. 

Let us define the fractional phase shift change on the GW phase introduced by $\hat{\alpha}_j$ 
by
\begin{equation}
    \epsilon_\psi \equiv \frac{\psi_{3j+8}(f_d)-\psi_{3j+8,{\rm GR}}(f_d)}{\psi_{3j+8,{\rm GR}}(f_d)}.
        \label{eq:eps_psi}
\end{equation}
(See Eqs.~\eqref{eq:GW_dep}-\eqref{eq:GW_dep2}.)
In addition, for this study, we consider the additional constraint on the speed of gravity set from GW170817 \cite{2017ApJ...848L..13A} and GRB170817A which can be converted into a constraint on $\hat{\alpha}_j$, i.e.
\begin{equation}
    \left|\frac{\hat{\alpha}_j}{2} f^j_{R,d}\right|<10^{-15},
    \label{eq:cond_t}
\end{equation}
where $f_{R,d}$ is the GW170817 merger frequency, which is of the order of $f_s=2000$ Hz \cite{2019PhRvX...9a1001A}.

Fig.~\ref{fig:BNS_dphi_difference} shows the fractional phase shift introduced by the  $\hat{\alpha}_j$ for each dispersion relation (alongside the PN order to which they contribute). In each case the range of the $y$-axis satisfies (\ref{eq:cond_t}), and one observes that the fractional GW phaseshift introduced by $\hat{\alpha}_j$ varies over many orders of magnitude. In particular, for the largest values of $\hat{\alpha}_j$ allowed by (\ref{eq:cond_t}), the fractional GW phaseshifts are very large, of order $10^4$. In general the measurements of GW phaseshifts are more accurate \cite{2019PhRvD.100j4036A}, and hence one expects the PN measurements to give better constraints than the GW-GRB time delay.
For instance in the case of massive gravity (panel with $j=-2$ in Fig.~\ref{fig:BNS_dphi_difference}),
one can see that a mass of the graviton of the order of $2 \cdot 10^{-20}$ eV/$c^2$, emitted by a source at redshift $\sim 0.01$, would modify the corresponding PN coefficient of $10^{4}$ times its GR value.
For GW170817, the constraint on the GR value of this PN coefficient is of the order of $10 \%$ \cite{2019PhRvL.123a1102A}, which from Figure \ref{fig:BNS_dt_difference} would lead to an upper limit on the mass of the graviton of order a few times $10^{-22}$eV/$c^2$.

We now consider the fractional deviation introduced by the GW dispersion relation $\hat{\alpha}_j$ on the GW-EM time delay in Eq.~\eqref{eq:GWGRBtimedelay}. 
We define the time delay introduced by $\hat{\alpha}_j$ as
\begin{equation}
    \epsilon_t \equiv \frac{f_{R,d}^j}{2}\mathcal{T}_j (z,H_0,\hat{\alpha}_j).
    \label{eq:eps_t}
\end{equation}
The above quantity is measured in seconds since we have chosen for this definition $\Delta t_s^{\rm GW-EM}=0$.
Figs.~\ref{fig:BNS_dt_difference} shows the values of $\epsilon_t$ for the 8 dispersion relations corresponding to the PN orders that we can study from the GW phase.
Note that almost all the dispersion relations predicts the same maximum value of $\epsilon_t \sim 100s$.
The reason is that we are computing $\epsilon_t$ from the maximum value given in Eq.~\eqref{eq:cond_t}. 
As an example, consider the case of massive gravity, given by the dispersion relation with $j=-2$.  
From Fig.~\ref{fig:BNS_dt_difference} we see that one would need a massive graviton with mass of about $10^{-21}$ eV$/c^2$ for a GW170817 like-source  (redshift $\sim$ 0.01) in order to explain the 1.74 second delay observed (assuming that the GW and GRB were prompt at the same time).  
This limit is higher than those set in \cite{2019PhRvD.100j4036A}, which make use of the BBHs detections, and use the phase of study of the PN orders of the GW phase. 
We conclude that, with the current accuracy, the GW-EM time delay cannot constrain massive gravity better than the GW phase study.

For each GW dispersion relation model, one can predict which observable --- the GW phase or GW-GRB delay --- will best constrain the GW dispersion relation. 
The ratio $\epsilon_\psi/{\epsilon_t}$ (see Eqs.~\eqref{eq:eps_psi} and \eqref{eq:eps_t}) quantifies which of the two observables is more modified by the introduction of $\hat{\alpha}_j$. 
Assuming that our GW waveform model is well approximated by the PN expansion in Eq.~\eqref{eq:PNexp} and the reference GW frequency for computing the GRB time delay is the last stable orbit $f_{R,d}=(6^{3/2} \pi M)^{-1}$, we obtain
\begin{eqnarray}
    && \frac{\epsilon_\psi}{\epsilon_t}=
     \frac{\pi \mathcal{T}_j}{(j+1)\beta_{3j+8}^{\rm PN}}\frac{2}{f_{R,d}^{j} \mathcal{T}_j} =  \frac{256 \nu 6^{\frac{3}{2}j}}{3(j+1)M g_{3j+8}(\nu,S_1,S_2)}  \nonumber \\ &&
    \qquad \qquad \qquad \qquad \qquad \qquad \qquad   \text{ when} \, j\neq -1
     \\
     &&
     \frac{\epsilon_\psi}{\epsilon_t}=
     \frac{\pi \mathcal{T}_{-1}}{\beta_{5,\ln}^{\rm PN}}\frac{2}{f_{R,d}^{-1} \mathcal{T}_{-1}} =  \frac{256 \nu 6^{-\frac{3}{2}}}{3 M g_{5,{\rm \ln}}(\nu,S_1,S_2)} \nonumber \\ &&
    \qquad \qquad \qquad \qquad \qquad \qquad \qquad   \text{ when} \, j= -1
\end{eqnarray}

It is important to observe that in these ratios, all dependence on the cosmological parameters has dropped out, as has the dispersion relation parameter $\hat{\alpha}_j$. 
Indeed the ratios depends only on the source masses and spins. 
In Fig.~\ref{fig:phi_accuracy} we show $\epsilon_\psi/{\epsilon_t}$ computed for several values of the binary masses (with spins fixed to 0).
As anticipated in the previous paragraphs, for all the GW dispersion relations the changes introduced by the GW phase are order of magnitudes larger than those introduced in the GW-GRB time delay.
For instance, in the case of massive gravity($j=-2$),as in the majority of the other dispersion relations, for a $1.4-1.4 M_{\odot}$, the GW phase is modified $\sim 10^4$ times more that the GW-EM time-delay.
This means that unless we are able to measure the GW-GRB delay with precision $10^4$ times better than the GW phase, most of the information on the parameter $\hat{\alpha}_j$ will come from the GW phase alone.
However, if the GW-EM time delay is constrained $10^4$ times better than the GW phase, then a more stringent upperlimits on the GW dispersion relation will come from the GW-GRB time delay.
For BNSs, the fractional deviation on the PN coefficients with $n=4,6$ and $7$ is currently measured with an accuracy of 200-500\% \cite{2019PhRvL.123a1102A}. This means that, if we observe a GW-GRB time delay of $\sim 1$~s,  we would need an accuracy of the order of the millisecond to improve to the constraint on the corresponding GW dispersion relations.
We will see how this prediction valid in Sec. \ref{sec:5}.

\section{A Complete inferencial method\label{sec:4}}

In the previous section, in order to gain some intuition, we fixed the value of the Hubble constant $H_0$. However, since one of our aims is to probe the Hubble constant tension \cite{2017NatAs...1E.169F}, we now consider $H_0$ to be unknown.

Notice that the Hubble constant $H_0$ appears in each observable we have defined, see Eqs.~\eqref{eq:GWGRBtimedelay}, \eqref{eq:GW_lum} and \eqref{eq:GW_dep}.
In particular, since $H_0$ appears in the denominator of each of these observables, it follows that decreasing the value of $H_0$ will result in a stronger phase shift and longer GW-GRB delay (if $\hat{\alpha}_j \neq 0$) and in a higher luminosity distance for the GW. 
On the other hand, increasing $H_0$, will result in a smaller phase shift and shorter GW-GRB time delay  and in a lower luminosity distance for the GW.
It is then clear that the measurement of $\alpha_M$ and $\hat{\alpha}_j$ will correlate through our current uncertainties on $H_0$. This is indeed a crucial point. It means for instance that if one fixes the value of $H_0$ to a slightly different value, a biased value of $\alpha_M$ and $\hat{\alpha}_j$ will be recovered.
 
In this section we present a statistical method able to take all these effects into account.
Let us discuss which are the statistical variables of the model. 
We assume a FLRW background specified by some parameters $\Lambda$ (in the particular case we consider here $\Lambda$ is simply the parameter $H_0$), a deviation in the GW friction term $\alpha_M$ and dispersion relation encoded in $\hat{\alpha}_j$. These are the {\it population parameters} on which we would like have posterior distributions. 
Furthermore, there are other parameters which are intrinsic to each source but not measured or provided as posteriors: in particular, the initial prompt-delay between the GW and its EM counterpart $\Delta t_s$ and the merger frequency at the detector $f_d$.
These, together with $H_0$ and the source redshift\footnote{For the sake of notation, here $z$ is the cosmological redshift obtained after the correction for peculiar velocities} $z$ will determine  values for the three observables: the GW luminosity distance $d_{\rm GW}$, the GW-EM time delay $\Delta t_d$ and the GW phase shift $\delta \psi$.
These observables are measured from several observed datasets. The GW luminosity distance and phase shift are measured from GW data $x_{\rm GW}$, the redshift is observed from the hosting galaxy observations $x_{z}$ and the GW-GRB time delay is observed from GRB data $x_{\rm EM}$. 
Let us refer to these three datasets as $\vec{x}$.
\begin{figure}[ht!]
    \centering
    \includegraphics[scale=0.75]{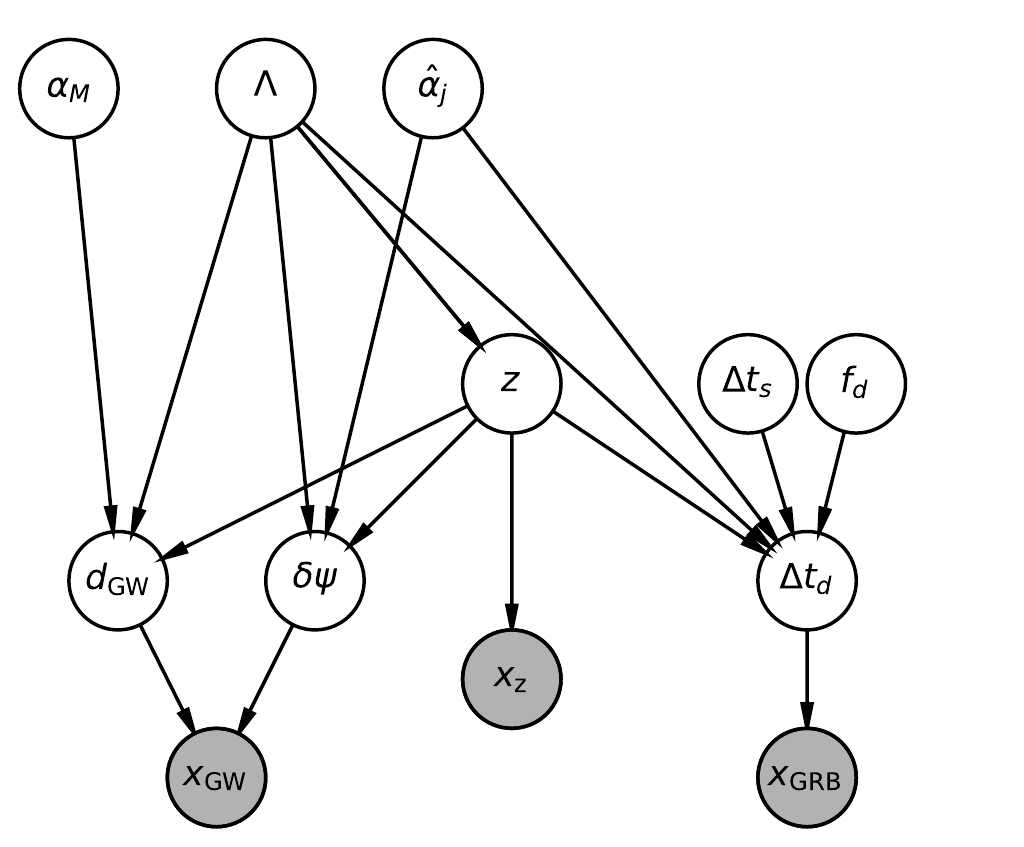}
    \caption{The inferencial model. The arrows between the nodes represent conditional probabilities between the variables, while the shaded nodes represent observed variables.}
    \label{fig:bayesian_model_complete}
\end{figure}

The inferencial method is represented in Fig.~\ref{fig:bayesian_model_complete} and encodes all the conditional dependencies between our variables \cite{2020arXiv200200269H}.
From the Bayesian network, it is immediate to see that if $\Lambda$ and redshift are given (fixed to a value), then $\alpha_M$ and $\hat{\alpha}_j$ are conditionally independent, i.e. $p(\alpha_M, \hat{\alpha}_j|\Lambda,z)=p(\alpha_M|\Lambda,z)p( \hat{\alpha}_j|\Lambda,z)$. This means that if we are provided with a perfect knowledge of the Hubble constant and the source redshift, then GW friction and dispersion can be measured independently from each other. However, this is not true if we want to measure also the Hubble constant.

In our analysis we are interested in the likelihood of observing the three datasets $x_{\rm GW}$, $x_{z}$ and $x_{\rm EM}$ given some values of the FLRW background parameters and the deviation parameters $\alpha_M$ and $\hat{\alpha}_j$.
We would like to sample the posterior probability $p(\lambda,\alpha_M,\hat{\alpha}_j|\vec{x})$. This is given by

\begin{equation}
    p(\Lambda,\alpha_M,\hat{\alpha}_j|\vec{x})=\frac{p(\Lambda,\alpha_M,\hat{\alpha}_j ,\vec{x})}{p(\vec{x})}.
    \label{eq:rough_likelihood}
\end{equation}
Let us focus on the numerator in the above Eq., this can be factorized following the Fig.~\ref{fig:bayesian_model_complete}:
\begin{eqnarray}
    &p(\vec{x},\Lambda,\alpha_M,\hat{\alpha}_j)=\int p(\alpha_M)p(\Lambda)p(\hat{\alpha}_j)p(\Delta t_s)p(f_d)p(z|\Lambda) \cdot \nonumber\\& \cdot p(d_{\rm GW}|\alpha_M,\Lambda,z)p(\delta \psi|\hat{\alpha}_j,\Lambda,z)p(\Delta t_d|\hat{\alpha}_j,\Lambda,\Delta t_s,f_d) \cdot \nonumber \\
    & \cdot p(x_{\rm GW}|d_{\rm GW},\delta \psi) p(x_{\rm z}|z) p(x_{\rm EM}|\Delta_d) \nonumber
        \label{eq:full_likelihood}
\end{eqnarray}
In Eq.~\eqref{eq:full_likelihood}, the marginalization is carried out on all the variables except for $\Lambda,\alpha_M,\hat{\alpha}_j$.
In Eq.~\eqref{eq:full_likelihood}, the terms $p(\alpha_M),p(\Lambda),p(\hat{\alpha}_j)$ represent the priors probabilities. 
The terms $p(\Delta t_s),p(f_d)$ are prior distributions.
The probability $p(z|\Lambda)$ is a prior probability for the source to be located at a redshift $z$. We can chose for example a uniform in comoving volume prior, which in the local universe scales as $p(z|H_0)\propto z^2 /H_0^3$.
The probabilities $p(d_{\rm GW}|\alpha_M,\Lambda,z),p(\delta \psi|\hat{\alpha}_j,\Lambda,z),p(\Delta t_d|\hat{\alpha}_j,\Lambda,\Delta t_s,f_d)$ represents the probability of having a value for one of the three observables given some values for the FLRW background parameters and GR deviation parameters.
We assume these probabilities to be given by Dirac delta functions, since the observables can be computed deterministically from the relations that we have defined in Sec.~\ref{sec:2}.

With the above considerations Eq.~\eqref{eq:full_likelihood} reduces to
\begin{eqnarray}
    &p(\vec{x},\Lambda,\alpha_M,\hat{\alpha}_j)=p(\alpha_M)p(\Lambda)p(\hat{\alpha}_j) \int p(\Delta t_s)p(f_d)p(z|\Lambda) \cdot \nonumber\\ & \cdot p(x_{\rm GW}|d_{\rm GW}(\alpha_M,\Lambda,z),\delta \psi (\hat{\alpha}_j,\Lambda,z)) p(x_{\rm z}|z) \nonumber \cdot \\ & \cdot p(x_{\rm EM}|\Delta t_d (\hat{\alpha}_j,\Lambda,\Delta t_s,f_d,z)) dz d f_d d\Delta t_s 
    \label{eq:full_likelihood2}
\end{eqnarray}
The remaining terms  in Eq.~\eqref{eq:full_likelihood2} are the likelihoods, which are computed from the different datasets. 
Often we will not have access to the likelihood values but to the posterior distributions. 
For instance, we will have access to the joint posterior for the GW luminosity distance and PN parameters. 
The posteriors can be used in our method using the Bayes theorem
\begin{equation}
   p( x_{\rm GW}| d_{\rm GW},\delta \psi)= \frac{p(d_{\rm GW},\delta \psi | x_{\rm GW}) p(x_{\rm GW})}{\pi(d_{\rm GW},\delta \psi)},
   \label{eq:bayes}
\end{equation}
where $\pi(\cdot)$ is the prior used to generate the posterior samples. Therefore, we can compute the posterior $p(\Lambda,\alpha_M,\hat{\alpha}_j|\vec{x})$ using Eq.~\eqref{eq:bayes}  and the Bayes theorem as
\begin{eqnarray}
&p(\vec{x},\Lambda,\alpha_M,\hat{\alpha}_j)=p(\alpha_M)p(\Lambda)p(\hat{\alpha}_j) \int p(\Delta t_s)p(f_d)p(z|\Lambda) \cdot \nonumber\\ & \cdot p(\vec{x}) \frac{p(d_{\rm GW}(\alpha_M,\Lambda,z),\delta \psi (\hat{\alpha}_j,\Lambda,z)|x_{\rm GW})}{\pi(d_{\rm GW}(\alpha_M,\Lambda,z),\delta \psi (\hat{\alpha}_j,\Lambda,z))} \frac{p(z|x_{\rm z})}{\pi(z)} \nonumber \cdot \\ & \cdot \frac{p(\Delta t_d (\hat{\alpha}_j,\Lambda,\Delta t_s,f_d,z) |x_{\rm EM})}{\pi(\Delta t_d (\hat{\alpha}_j,\Lambda,\Delta t_s,f_d,z))} dz d f_d d\Delta t_s.
\label{eq:final_likelihood}
\end{eqnarray}
Finally, we can plug the joint probability in Eq.~\eqref{eq:final_likelihood} in Eq.~\eqref{eq:rough_likelihood} to obtain
\begin{eqnarray}
&p(\Lambda,\alpha_M,\hat{\alpha}_j|\vec{x})=p(\alpha_M)p(\Lambda)p(\hat{\alpha}_j) \int p(\Delta t_s)p(f_d)p(z|\Lambda) \cdot \nonumber\\ & \cdot \frac{p(d_{\rm GW}(\alpha_M,\Lambda,z),\delta \psi (\hat{\alpha}_j,\Lambda,z)|x_{\rm GW})}{\pi(d_{\rm GW}(\alpha_M,\Lambda,z),\delta \psi (\hat{\alpha}_j,\Lambda,z))} \frac{p(z|x_{\rm z})}{\pi(z)} \nonumber \cdot \\ & \cdot \frac{p(\Delta t_d (\hat{\alpha}_j,\Lambda,\Delta t_s,f_d,z) |x_{\rm EM})}{\pi(\Delta t_d (\hat{\alpha}_j,\Lambda,\Delta t_s,f_d,z))} dz d f_d d\Delta t_s.
\label{eq:final_likelihood2}
\end{eqnarray}

The posterior distribution in Eq.~\eqref{eq:final_likelihood} does not consider yet selection effects \cite{2020ApJ...891L..31F,2019MNRAS.486.1086M}, which takes into account that for some values of $\Lambda,\alpha_M,\hat{\alpha}_j$ some binaries are more probable to be detected. 
When combining multiple sources for inferring the population parameters $\Lambda,\alpha_M,\hat{\alpha}_j$ this is a crucial term to include in order to recover a measure which is unbiased by the selection bias.
Selection effects can be included dividing Eq.~\eqref{eq:final_likelihood2} by the selection probability 
\begin{eqnarray}
&\beta(\Lambda,\alpha_M,\hat{\alpha}_j)=\int P_{\rm det}^{\rm GW}(z,\Lambda,\alpha_M,\hat{\alpha}_j)  P_{\rm det}^{\rm z}(z)  p(z|\Lambda) \nonumber \\ & P_{\rm det}^{\rm GRB}(z,\Lambda,\hat{\alpha}_j,f_d,\Delta t_s) p(f_d)p(\Delta t_s) dz df_d d\Delta t_s.
\label{eq:selef}
\end{eqnarray}
In the above Eq., $P_{\rm det}^{\rm GRB-GW-z}$ are the detection probabilities of the different datasets given the parameters of the binary.
It is usually assumed that the detection probabilities will be dominated by the GW detection probability\cite{2018Natur.562..545C} as it will go to zero faster with respect to the other detection probabilities. 
In principle the GW detection probability will also depend on the phase shift of the GW, but at small redshift and small values of $\hat{\alpha}_j$ it is usually assumed that the GWs modified by a non-GR dispersion relation can be always detected \cite{1998PhRvD..57.2061W}.

\section{Simulations \label{sec:5}}

In this section we implement the statistical framework discussed previously showing its statistical properties with software injections.
Let us consider a Universe with $H_0=70 {{\rm\, km\, Mpc}^{-1}s^{-1}}$, $\alpha_M=5$ and $\hat{\alpha}_{-2}=9.23 \cdot 10^{-14} \rm{Hz}^2$ (which corresponds to a massive graviton of $m_g=2 \cdot 10^{-22} {\rm eV}/c^2$).
We simulate 100 BNSs mergers in a Universe with a uniform in comoving volume redshift prior 
\begin{equation}
    p(z|H_0)\propto \frac{1}{E(z)H_0^3} \left[ \int_0^z \frac{dz'}{E(z')} \right]^2.
\end{equation}
We consider that GWs with a luminosity distance $<100$ Mpc are always detected. 
This choice is motivated from a GW170817-like scenario, with SNR of $33$ at $40$ Mpc that would correspond to an SNR of $12$ at $100$ Mpc. Note that this assumption is only used to decide if a GW event is detected, and it is used to evaluate the selection effect in Eq. \eqref{eq:selef}.
We also assume that for each event we perfectly know the sky-location, the redshift and the merger frequency (which we fix for each event to 2000 Hz). We assume that the emission delay between GW-GRB, $\Delta t_s$, is generated uniformly between 0 and -10 seconds, following the model assumptions in \cite{2017ApJ...848L..13A}.

For each BNS event we generate the recovered posterior samples for the GW luminosity distance, phase shift and GW-GRB delay using the following assumptions.
We assume the posterior $p(\delta \psi,d_{\rm GW}|x_{\rm GW})$ as a multi-variate gaussian distribution centered around the signal values.
Following \cite{2018Natur.562..545C}, the standard deviation of the luminosity distance is assumed to be
\begin{equation}
    \sigma_d=(1.8/ \rho) d_{\rm{GW}},    
\end{equation}
of the injected distance value. 
Regarding the phase shift, we assume a standard deviation of the $10\%$ its injected value\cite{2019PhRvL.123a1102A}.
For the detected GW-GRB time delay we assume an uncertainty of $0.05s$ on its injected value. The choice of these accuracies will enter the numerator of Eq.~\eqref{eq:final_likelihood} and drive the uncertainties on the population parameters.

Let us first show that when we fix the value of $H_0$, the two GR deviation parameters $\alpha_M$ and $\aj$ are independent each other, as predicted by the statistical framework.
As an example, we show the posterior samples of a single BNS at redshift $z=0.010$, in our simulated Universe this event have a luminosity distance of $d_{\rm GW}=48$ Mpc and SNR of $\rho=27.5$.

\begin{figure}[htp!]
    \centering
    \includegraphics[scale=0.5]{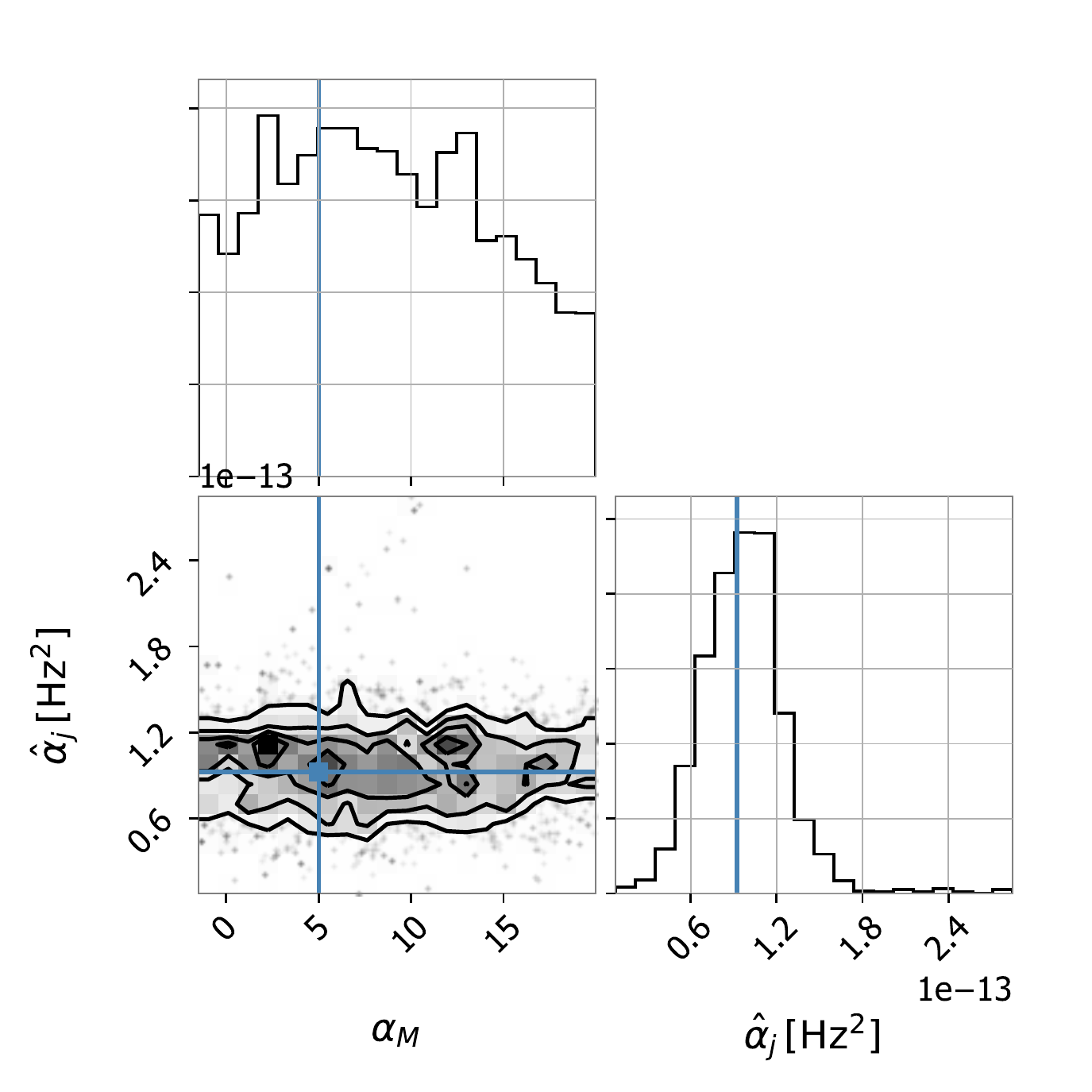}
    \caption{Marginal and joint posterior distributions for $\alpha_M, \hat{\alpha}_{-2}$ for a software injection with SNR 27 in modified cosmology.}
    \label{fig:single_inj_am0_at0}
\end{figure}

By fixing $H_0$ to the injected value, we obtain $\alpha_M=-8.20^{+6.10}_{-6.84e}$,$\hat{\alpha}_{-2}=9.68^{+2.40}_{-2.77} \cdot 10^{-14} {\rm Hz}^2$ (median with $68\%$ confidence level).
Fig.~\ref{fig:single_inj_am0_at0} shows the joint and marginal posterior distributions for $\alpha_M$ and $\hat{\alpha}_{-2}$.
The two deviation parameters for GR are not crosscorrelated. This is due to the fact that the value of $H_0$ is fixed.

\begin{figure}[h!]
    \centering
    \includegraphics[scale=0.35]{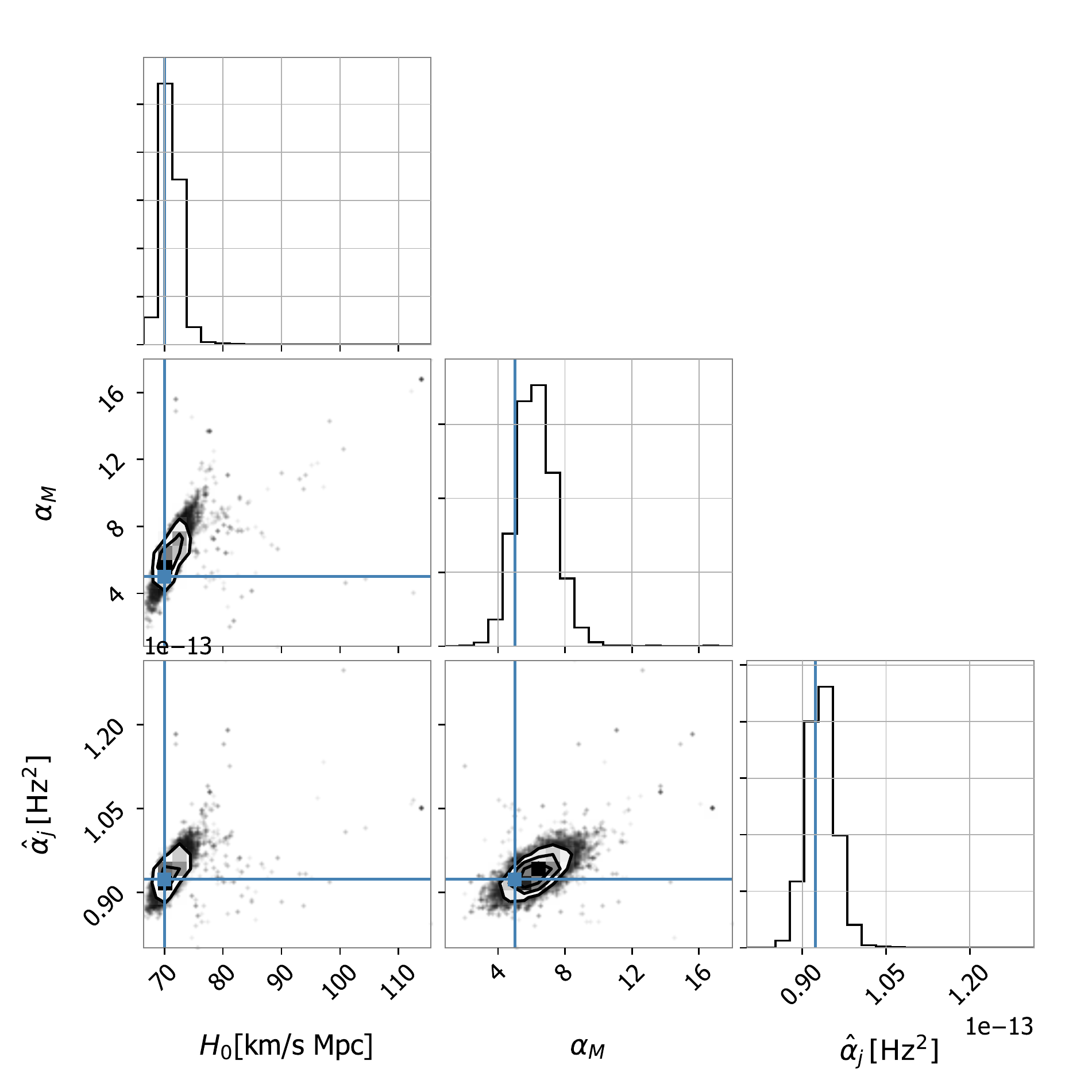}
    \caption{Marginal and joint posterior distributions 100 BNS detection in a modified Universe with parameters  $H_0=70{{\rm\, km\, Mpc}^{-1}s^{-1}}$, $\alpha_M=5$ and $\hat{\alpha}_{-2}=9.23 \cdot 10^{-14} \rm{Hz}^2$.}
    \label{fig:combination}
\end{figure}

The previous statement, however, it is not true if we consider $H_0$ unknown, and especially if we perform a population analysis.
In Fig.~\ref{fig:combination} we show the combined results on $H_0,\alpha_M$ and $\alpha_{-2}$ for the 100 simulated signals. 
It is possible to see that the posterior distributions reached the gaussian convergence, as expected when combining a large number of events.
However, it is important to note that even though the posterior has reached gaussian convergence, the cosmological parameters are still correlated and the final distribution on $H_0, \alpha_M$ and $\alpha_j$ is indeed a multivariate gaussian distribution with non diagonal covariance matrix. 
It is possible to observe that $\alpha_M$ and $\hat{\alpha}_j$ are positive correlated due to the presence of $H_0$.
The recovered values are $H_0=71.0^{+1.6}_{-1.4} {{\rm\, km\, Mpc}^{-1}s^{-1}}$ $\alpha_M=6.22^{+1.16}_{-1.08}$ and $\hat{\alpha}_{-2}=9.33^{+0.25}_{-0.25} \cdot 10^{-14} {\rm Hz^2}$, which corresponds to and accuracy of 2 \%, 15\% and 2 \%.
In other words, any observation aiming at constraining $\alpha_M$ and $\hat{\alpha_j}$ to a level of 15\% and 2 \% should also take into account a varying $H_0$ if our lack of knowledge on this parameter is higher than the $2\%$.
Currently, the tension on the Hubble constant is of the order of 6\% between the CMB and local universe observations. 
Therefore, we cannot reach a 15\% and 2\% accuracy on $\alpha_M$ nor $\hat{\alpha}_j$ by fixing $H_0$ to the CMB or local Universe value.

\section{Application to GW170817 \label{sec:GW170817}}
We now apply our framework to the case of GW170817 and its EM counterpart. In this section we present our results for this event, and their implications for cosmology and modified theories of gravity.

Using the statistical described in Sec.\ref{sec:4}, we calculate the joint posterior on $H_0$, $\alpha_M$, $\hat{\alpha}_j$ and $\Delta t_s$, given the observations of GW170817 together with its associated Gamma-ray Burst
and hosting galaxy NGC4993.
On denoting the LVC data by $x_{\rm GW}$, the GRB data by $x_{\rm EM}$ and the redshift by $x_{\rm z}$, and the combined measurements provided by these 3 datasets by $\vec{x}$, the posterior can be written using Eq.~\eqref{eq:final_likelihood2}. It is given by
\begin{eqnarray}
&&p(H_0,\alpha_M,\hat{\alpha}_j,\Delta t_s|\vec{x}) = \frac{p(\alpha_M)p(H_0)p(\hat{\alpha}_j) p(\Delta t_s)}{\beta(H_0,\alpha_M)} \nonumber \\ &&  \int dz d f_{R,d} \;
p(z|H_0)  p(f_{R,d}) \times \nonumber  \\ &&  \times \frac{p(d_{\rm GW},\delta \psi|x_{\rm GW})}{\pi(d_{\rm GW},\delta \psi)}  \frac{p(z|x_{\rm z})}{\pi(z)}  \frac{p(\Delta t_d  |x_{\rm EM})}{\pi(\Delta t_d)}  
\label{eq:pos}
\end{eqnarray}
where, as above, the functions $\pi(\cdot)$ are the priors used by the independent measurements to generate the posterior distributions. 
For LVC data, posteriors are generated with a uniform prior on the PN deviations and a $d^2_{\rm GW}$ prior on the GW luminosity distance.
For NGC4993 and the GW-GRB time delay, we assume that the posterior values are originally generated with a uniform prior. 
The selection effect is calculated using a Monte Carlo study with LIGO and Virgo sensitivity curves at the epoch of GW170817, more details  are given in Appendix \ref{app:A}.
We sample from the joint posterior distribution in Eq.~\ref{eq:pos} using a Monte Carlo Markov chain algorithm which details are given in Appendix \ref{app:B}.

\subsection{Priors and data used}
For the merger frequency $f_{d}$, we assume a uniform prior distribution between $2200$~Hz and $3500$ Hz, which are the minimum and maximum values found with different GW waveform models \cite{2019PhRvX...9a1001A}.
As in the analysis analysis in \cite{2017ApJ...848L..13A}, for the GW-GRB emission delay, we use a uniform prior $p(\Delta t_s)$ between $[-10,0]$s (we allow the GRB  emitted up to $10$~s after the GW at the source). 
For $\hat{\alpha}_j$ we use a uniform prior in $[\hat{\alpha}_{j,{\rm min}}, \hat{\alpha}_{j,{\rm max}}]$, where these two boundaries are computed from the condition $|\hat{\alpha}_j f_{R,d,{\rm min}}^j| <10^{-15}$ with  $f_{R,d,{\rm min}}=2200$~Hz.
This condition is set to satisfy the model independent constraints on the GW speed in \cite{2017ApJ...848L..13A} (also computed allowing the GRB to be emitted 10 seconds after the GW).
For the Hubble constant we use a flat in log prior between $[40, 110] \hu$ and for the parameter $\alpha_M$ a uniform prior between $[-10, 90]$.
The lower limit of the $\alpha_M$ prior has been chosen such that $d_{\rm GW}$ is a monotonically increasing function up to redshift $z=0.2$ following Eq.~\eqref{eq:al_con}, which clearly includes the redshift observed for GW170817.
The prior $p(z|H_0)$ for the cosmological redshift given the Hubble constant value is uniform in comoving volume.

\begin{table*}[t!]
\caption{Median and $1\sigma$ confidence intervals for $H_0,\aj$ (3$rd$, 4$th$ and 5$th$ columns) for \textit{Scenarios I-II}. For $H_0$ and $\alpha_M$ in \textit{Scenario II} we report the 95\% CL lower and upper bounds (due to their degeneracy described in the text).
First and Second column: $v_g$ scaling and correspondent PN for the GW dispersion relations. Correlations between variables are reported in the $6th,\, 7th,\, 8th$ columns.}
\begin{tabular}{||>{\centering}p{10mm}|>{\centering}p{10mm}|>{\centering}p{15mm}|>{\centering}p{15mm}|>{\centering}p{26mm}|>{\centering}p{26mm}|>{\centering}p{10mm}|>{\centering}p{9mm}|>{\centering}p{9mm}|>{\centering}p{12mm}|>{\centering\arraybackslash}p{12mm}|}
\hline
{\centering}
$v_g$&PN & \multicolumn{2}{c|}{$H_0 [\hu$]} &\multicolumn{2}{c|}{$\aj[{\rm eV}^{-j}c^{2j}]$}& $\alpha_M$ & \multicolumn{2}{c|}{$\mathcal{C}_{H_0-\aj}$} & $\mathcal{C}_{H_0-\alpha_M}$ & $\mathcal{C}_{\aj-\alpha_M}$ \\
\textit{I-II}&\textit{I-II} &\textit{I}&\textit{II}&\textit{I}&\textit{II} &\textit{II}& \textit{I}&\textit{II} &\textit{II}&\textit{II}\\
\hline \hline
$f_{d}^{-8/3}$ & 0 & $75^{+15}_{-9}$& $>63$ & $-8.5^{+6.7}_{-6.6} \cdot 10^{-52}$ & $-9.9^{+7.6}_{-8.9} \cdot 10^{-52}$  &$<54$ & $0.11$ & $-0.07$ & $0.64$ & $-0.31$\\

\hline
$f_{d}^{-7/3}$ & 1 & $75^{+14}_{-8}$& $>64$ & $6.7^{+9.1}_{-7.4} \cdot 10^{-49}$ & $7.8^{+10.8}_{-8.4} \cdot 10^{-49}$ & $<50$ & $0.09$ & $0.15$ & $0.72$ & $0.15$\\

\hline
$f_{d}^{-2}$ & 2 & $75^{+15}_{-8}$& $>64$ & $1.8^{+2.5}_{-1.9} \cdot 10^{-44}$ & $2.0^{+3.0}_{-2.2} \cdot 10^{-44}$ & $<48$ & $0.03$ & $0.08$ & $0.73$ & $0.08$\\

\hline
$f_{d}^{-5/3}$ & 3 & $75^{+14}_{-8}$& $>64$ & $3.6^{+7.2}_{-6.7} \cdot 10^{-40}$ & $4.7^{+7.9}_{-7.2} \cdot 10^{-40}$ & $<48$ & $0.05$ & $0.11$ & $0.71$ & $0.12$\\

\hline
$f_{d}^{-4/3}$ & 4 & $76^{+14}_{-9}$& $>66$ & $-1.9^{+1.5}_{-2.0} \cdot 10^{-35}$ & $-2.1^{+1.8}_{-2.5} \cdot 10^{-35}$ & $<48$ & $-0.17$ & $-0.11$ & $0.68$ & $-0.08$\\

\hline
$f_{d}^{-1}$ & 5l & $77^{+15}_{-9}$& $>66$ & $-3.2^{+2.5}_{-2.9} \cdot 10^{-31}$ & $-3.8^{+2.9}_{-3.5} \cdot 10^{-31}$ & $<48$ & $-0.10$ & $-0.19$ & $0.70$ & $-0.18$\\

\hline
$f_{d}^{-2/3}$ & 6 & $76^{+15}_{-9}$& $>67$ & $3.7^{+11.9}_{-4.8} \cdot 10^{-28}$ & $4.4^{+13.4}_{-6.0} \cdot 10^{-28}$ & $<48$ & $0.10$ & $0.08$ & $0.66$ & $0.03$\\

\hline
$f_{d}^{-1/3}$ & 7 & $76^{+15}_{-9}$& $>65$ & $-2.5^{+1.8}_{-2.8} \cdot 10^{-22}$ & $-2.8^{+2.0}_{-3.1} \cdot 10^{-22}$ & $<48$ & $-0.18$ & $-0.19$ & $0.67$ & $-0.11$\\
\botrule
\end{tabular}
\label{tab:results_table}
\end{table*}

When assuming no GW dispersion relation ($\hat{\alpha}_j=0$), we compute $p(d_{\rm GW}|x_{\rm GW})$ using the ``high-spin'' posterior for $d_{\rm GW}$ provided by the LVC \footnote{https://dcc.ligo.org/LIGO-P1800061/public}  \cite{2019PhRvX...9a1001A} and generated by fixing the sky position to that of NGC4993 and using the GW waveform generator PhenomPNRT \cite{2014PhRvL.113o1101H}.
When we consider $\hat{\alpha}_j \neq 0$, we use the joint posterior samples\footnote{https://dcc.ligo.org/LIGO-P1800059-v8/public} on the GW luminosity distance and phase shift $p(d_{\rm GW}, \delta \psi | x_{\rm GW})$ from \cite{2019PhRvL.123a1102A}.
These are also generated using the PhenomPNRT \cite{2019PhRvD.100d4003D}. 
For $p(\Delta t_d | x_{\rm EM})$ we assume a Gaussian posterior with mean $-1.74s$ (the GW arrives earlier than the GRB) and standard deviation $0.05s$ \cite{2017ApJ...848L..13A}. 
For the posterior on NGC4993 redshift, we assume a bivariate gaussian distribution centered at the observed redshift value of $\hat{z}_{\rm obs}=0.011$ and peculiar motion redshift $\hat{z}_{\rm pec}=0.001\cdot 10^{-3}$  with standard deviations of $2 \cdot 10^{-4}$ and $5 \cdot 10^{-4}$ respectively \cite{2007ApJ...655..790C}.

\subsection{Results} 
We now discuss our measurements of $H_0$, $\alpha_M$ $\alpha_j$ and $\Delta t_s$ as well as their physical meaning.  First, however, before presenting our general results we carry out two checks: namely we show that
our framework reproduces the results on $H_0$ and $\alpha_M$ inferred from GW170817 in previous works. 

As a first check, we fix $\alpha_M=\aj=0$ (namely General Relativity is assumed to be the correct theory of gravity). We run over $H_0$ and obtain a value for the Hubble constant of $H_0=74^{+14}_{-7}\,\hu$, which is consistent with the value of \cite{2017Natur.551...85A,2019PhRvX...9a1001A}. 
Moreover, we obtain a GW-GRB emission delay of $\Delta t_s=-1.72^{+0.05}_{-0.05}s$, i.e.~the  GW-GRB observed delay is due to an initial one processed by the cosmological expansion. This is expected has no extra delay mechanisms are added by $\aj$.

\begin{figure}
    \centering
    \includegraphics[scale=0.6]{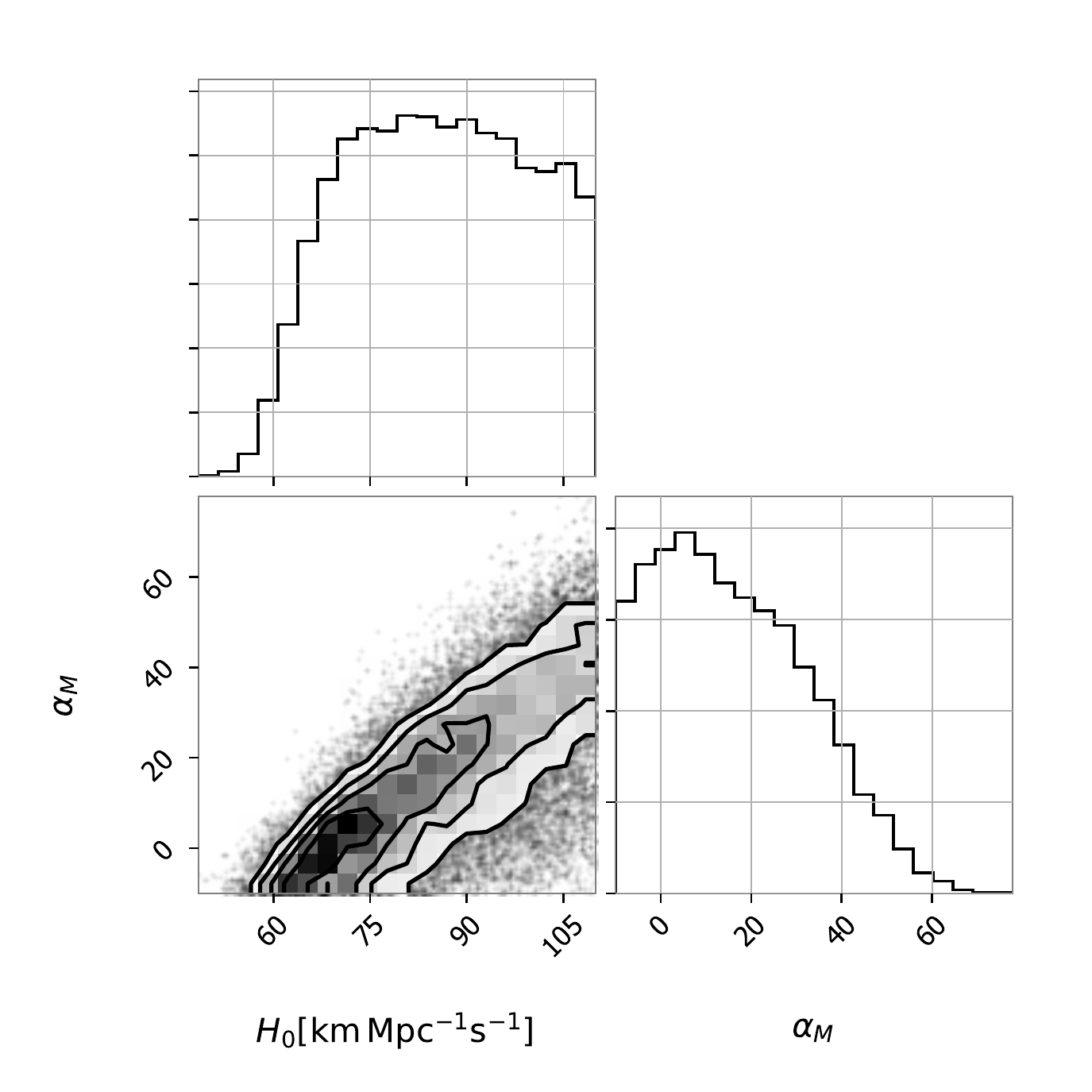}
    \caption{Joint and marginalized probability density function for $H_0$ and $\alpha_M$ assuming $\hat{\alpha}_j=0$ (no friction term).}
    \label{fig:fig1}
\end{figure}

In the second test case we consider a running of $H_0-\alpha_M$, a situation already analysed in \cite{2019PhRvD..99h3504L}, though using a different parametrization\footnote{In \cite{2019PhRvD..99h3504L} the GW friction is parameterized with a redshift independent parameter $c_M$ which can be related to $\alpha_M$ through $c_M = 2 \alpha_M$.} of $\alpha_M$.
The results we obtain, see Fig.~\ref{fig:fig1}, are in agreement with \cite{2019PhRvD..99h3504L} given the different prior choices.
Note that from Eq.~\eqref{eq:GW_lum} (giving the luminosity distance), the two parameters $\alpha_M$ and $H_0$ are degenerate. 
Hence with a single GW event, one can only provide constraints on one of the parameters, given a prior range on the other parameter.
Returning to the unique event GW170817, the constraints on $\alpha_M$ are computed given a chosen $H_0$ prior of $[40,110] \hu$, and we find $\alpha_M$ can be constrained $<46$ at 95\% CL.
Similarly, we can identify a lower bound for $H_0>63 \hu $ at 95\% CL, given by the fact that we are setting a theoretical cut-off on $\alpha_M>-10$ in our prior motivated by the condition that $d_{\rm GW}(z)$ must be an increasing function of redshift.
Also in this case, having no additional delay mechanism other than the cosmological expansion for the GW, also in this case, we obtain $\Delta t_s=-1.72^{+0.05}_{-0.05}s$.

We now apply our analysis to two Scenarios which are considered in this paper for the first time. In \textit{Scenario I} we consider several GW dispersion relations jointly with $H_0$ but fixing $\alpha_M=0$, while in \textit{Scenario II} we infer all the parameters (and allow also $\alpha_M$ to vary).

\subsubsection{Scenario I: $H_0 - \hat{\alpha}_j$}
\begin{figure}[h!]
	\hspace*{-0.3cm}                                          
    \includegraphics[scale=0.5]{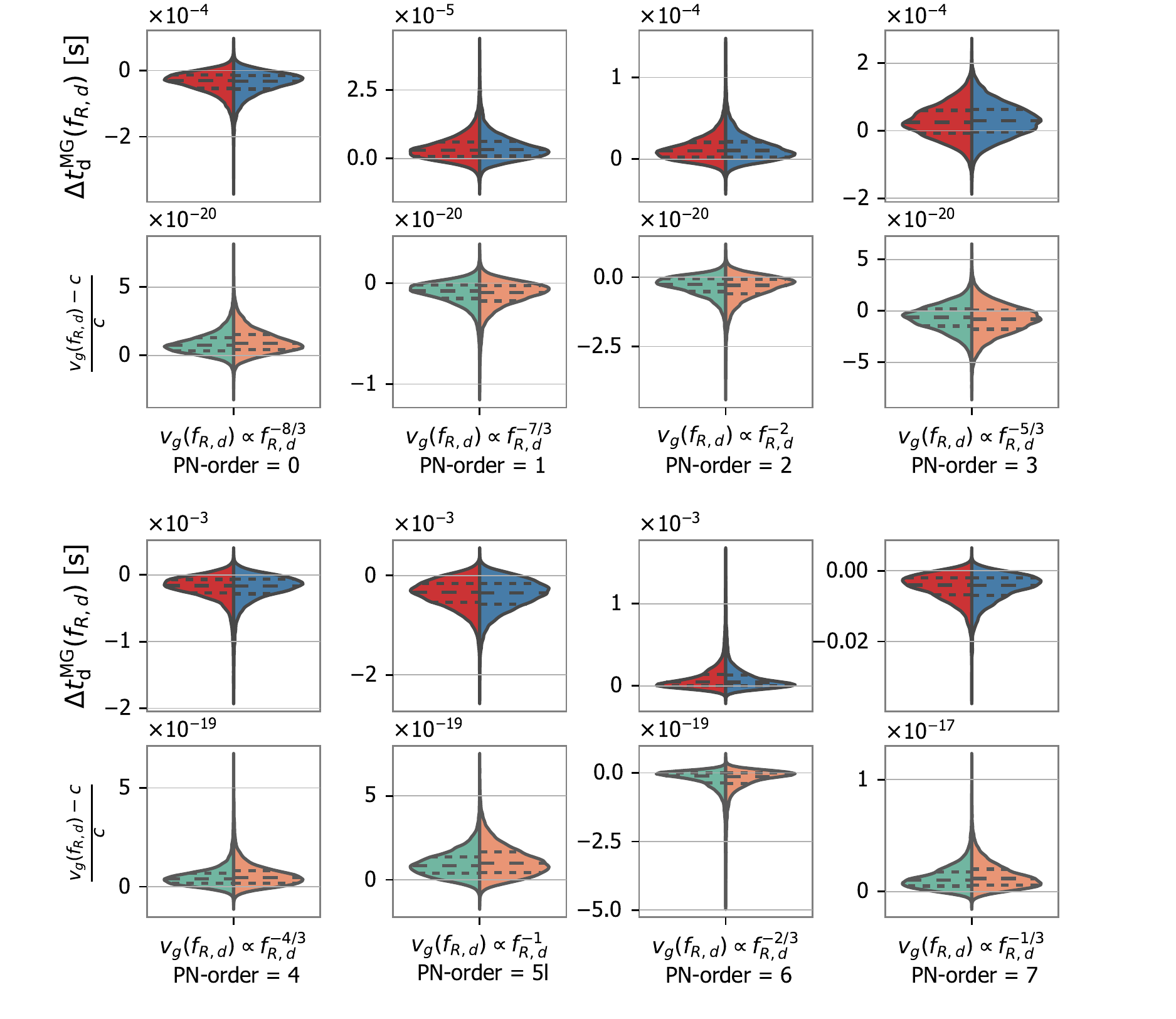}
    \caption{Posterior distributions for $\Delta t_d^{\rm MG}$ (GW-GRB observation delay) and $v_g/c-1$, both evaluated at $f_{R,d}$. Red/green: {\it scenario I}. Blue/orange: {\it scenario II}.}
    \label{fig:tg_vg}
\end{figure}

For {\it Scenario I}, we find that the results on the $H_0$ determination are mostly independent of the GW dispersion relation, see Tab.~\ref{tab:results_table}. 
This means that $H_0$ is mostly constrained through the GW luminosity distance and not the GW-GRB observation delay or GW phase.
Note that we considered both positive and negative values of $\hat{\alpha}_j$ in the posterior distributions, and thus include both sub- and super-luminal GW propagation. 
Indeed, here we take a purely phenomenological approach in order to fully exploit the information encoded in data.
It is important to observe that these combined observations also allow us to put constraints on the GW- GRB emission delay for several GW dispersion relations considered in this paper (see Fig.~\ref{fig:tg_vg}). 

As it can be seen, the contribution to the observation delay due to modification of gravity ($\Delta t_{d}^{\rm MG}=f_{R,d}^j \mathcal{T}_j/2$ in Eq.~\eqref{eq:GWGRBtimedelay}) is always negligible if compared to the current uncertainty on the observed time delay ($0.05$~s). 
As a consequence, for {\it all} the GW frequency dependent dispersion relations, the emission delay posterior converges to a value of $\Delta t_s=-1.72^{+0.05}_{-0.05}$s, and hence is consistent with the measured GW-GRB observation delay rescaled by a source redshift of $\sim 0.01$. 
It follows that the GW phase, i.e.~the deviation from the waveform PN parameters is setting a very tight constraint on the GW speed $v_g(f_{R,d})$ at the merger frequency with respect to the one set in \cite{2017ApJ...848L..13A} by the GW-GRB observation delay which considers a frequency independent GW dispersion relation ($j=0$).
Indeed, from a theoretical point of view one can not compare directly the results on the speed of gravity in this paper with the ones from \cite{2017ApJ...848L..13A} as we would be comparing frequency independent dispersion relations with a frequency dependent one.
However, from the results in this paper, one can conclude that for the frequency dependent dispersion relations that we can study with GW170817, only a negligible fraction of the GW-GRB observed time delay can be due to modified gravity.
Fig.~\ref{fig:tg_vg} also shows the posteriors for {\it scenarios I-II} obtained for the GW group velocity discrepancy from $c$.

\subsubsection{Scenario II: $H_0-\alpha_M-\hat{\alpha}_j$}

Finally, we consider the most general case with all the parameters, namely {\it scenario II}.  Due to the introduction of $\alpha_M$, the error budget for the  dispersion relations are $\sim$ 20\% worse than the previous case (see Tab.~\ref{tab:results_table}). 
The marginalization on $\alpha_M$ is contributing to the error budget of $\hat{\alpha}_j$.
In other words, for this kind of measurement, the GW friction and the GW dispersion relation are correlated with each other.
Indeed, correlations among the parameters that we want to infer play a crucial role for obtaining a bias-free measurement as we have seen in Sec.~\ref{sec:5}.

In Tab.~\ref{tab:results_table} we also report the correlations observed between the pairs $H_0,\alpha_M$ and $\hat{\alpha}_j$. We define the correlation as $\mathcal{C}=\rm{cov}(X,Y)/\sqrt{\rm{var}(X) \rm{var}(Y)}$, where ``cov'' and ``var''  are the covariance and variance operators.
In general for the pair $(H_0,\alpha_M)$  we observe a strong correlation. 
Indeed, we can obtain the same value of the posterior by increasing $\alpha_M$ and increasing $H_0$ or viceversa. 
This is consistent with the strong degeneracy of these two parameters in Eq.~\eqref{eq:GW_lum} for low redshift events. 
Concerning the pair $(H_0,\hat{\alpha}_j)$, for almost all the dispersion relations, we observe  weak correlation. 
This is mostly due to the fact that $H_0$ is being constrained from the GW luminosity distance together with $\alpha_M$.
Regarding the $(\alpha_M,\hat{\alpha}_j)$ correlation, this is not as significant as that of $(H_0,\alpha_M)$ but it is not completely negligible.
Indeed, when combining several events like GW170817 including the increase of the detector sensitivity, these correlations will become more and more important for the determination of $H_0, \alpha_M, \hat{\alpha}_j$ as we have shown in Sec.\ref{sec:5}.

\subsection{Implications}
Here we comment how the results of GW170817 can be understood in terms of different theories which modify gravity in  late-time universe. 
The following results are computed for {\it Scenario II} and hence refer to the general case in which we also allow a GW dispersion relation, and are reported with $3\sigma$ CL unless stated.

Regarding the GW dispersion relation, the factor $\hat{\alpha}_j$ can be linked to Lorentz invariance breaking \cite{2012PhRvD..85b4041M}, or  massive gravitons, or  scalar/vector fields coupled to the metric  \cite{2012PhRvD..86h4037B,2018FrASS...5...44E,2019RPPh...82h6901K}.
Many of these theories also have an $\alpha_M$ contribution, due for instance to a varying Planck mass.

\begin{figure*}[htp!]
    \includegraphics[scale=0.75]{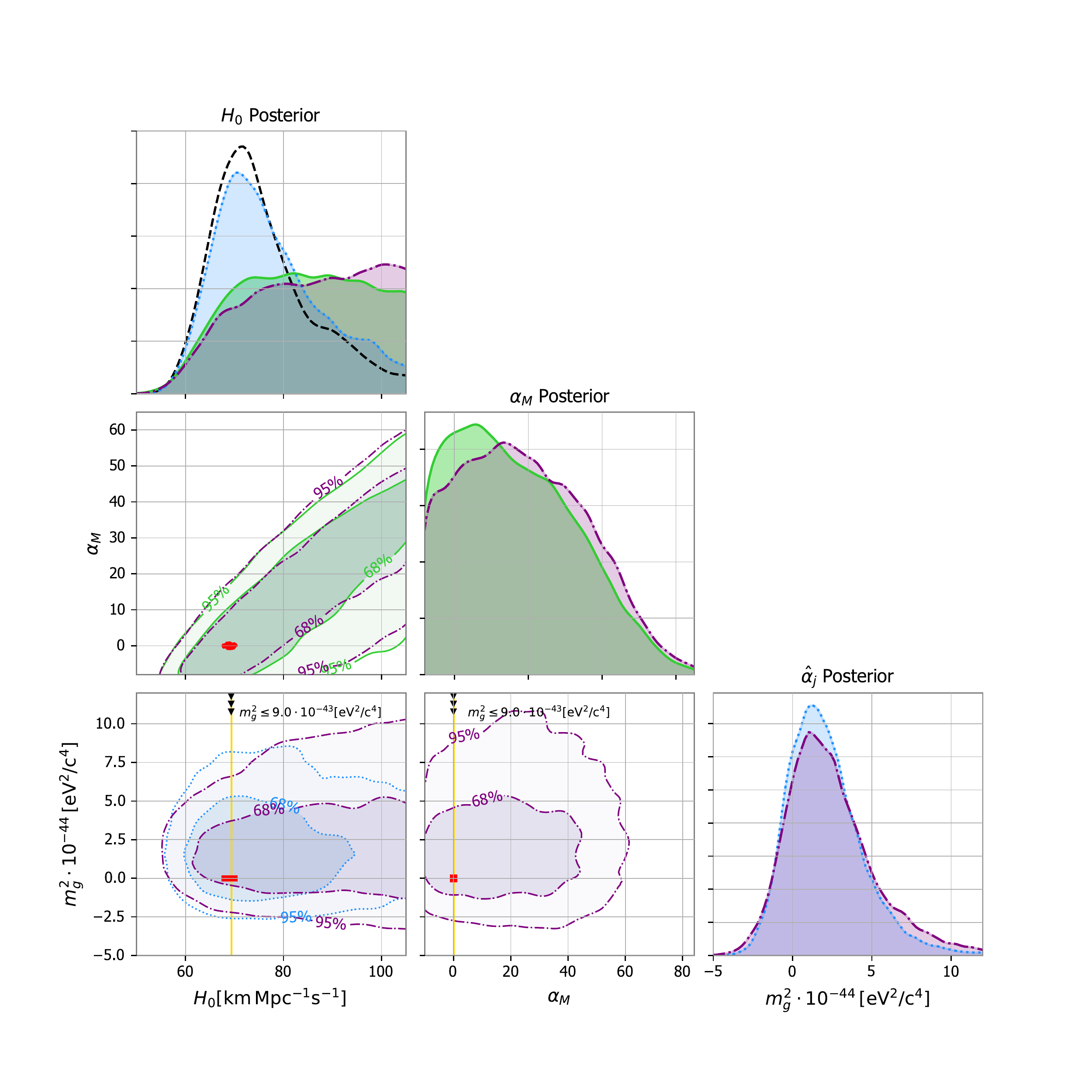}   
    \caption{Posterior distributions for $H_0,\alpha_M$ and $m_g^2$.  Black dashed line: $H_0$ only (namely GR). Green solid line: ($H_0,\alpha_M$). {\it Scenario I} ($H_0,\hat{\alpha}_j$): blue dotted line. {\it Scenario II} ($H_0,\alpha_M,\hat{\alpha}_j$): purple dash-dot line. Red patches: cosmological motivated constraints on $H_0,\alpha_M$ \cite{2019PhRvD..99j3502N} (fixing $m_g^2=0$). Yellow line: Upper bound on $m^2_g$ from GW170817 in \cite{2019PhRvL.123a1102A} (fixing $H_0=69.3 \hu$.)}
    \label{fig:gg}
\end{figure*}

For example, consider massive gravity (corresponding to $j=-2$), allowing also a non-zero $\alpha_M$.
The square of the graviton mass, $m_g^2$, is given by $\hat{\alpha}_{-2}$ divided by the square of the Planck constant.
Fig.~\ref{fig:gg} shows the posteriors on $H_0, \alpha_M$ and $m_g^2$.
In all the scenarios, the posteriors are compatible with GR at $1\sigma$~CL.
In the most general {\it scenario II}, we obtain  $m^2_g=2.0_{-5.8}^{+14.0} \cdot 10^{-44} \rm{eV^2/c^4}$. 
For comparison, the best indirect bounds for $m_g$ are provided from the solar system or weak lensing maps ($0 < m_g^2 \lesssim 10^{-64} \rm{eV^2/c^4}$) or  from binary pulsars ($0 < m_g^2 \lesssim 10^{-54} \rm{eV^2/c^4}$) \cite{2017RvMP...89b5004D}. 
This  constraint is tighter than the previous one using GW170817 ($m^2_g<9.0 \cdot 10^{-43} \rm{eV^2/c^4}$) \cite{2019PhRvL.123a1102A} and obtained by fixing $H_0$ and no information on redshift. 
The reason for this it is not due to the prior choices on any of our parameters (see below).\footnote{Note that in our approach, we use a flat prior on $\hat{\alpha}_{-2}$ and hence a flat prior on $m^2_g$. The correspondent prior on $m_g$ due to this choice would be $\propto m_g$, hence we prefer higher graviton masses in the analysis.}
Instead it is due to the PN parameters  posterior samples we are using in this analysis.
In fact, in \cite{2019PhRvL.123a1102A} a dedicated pipeline adding a GW dispersion relation on the entire waveform is used, while the posterior samples that we are using are generated with only the inspiral part (that for GW170817 gives the majority of the SNR). 
In order to show that the improvement is not due to any prior choice but on the PN posterior samples themself, we perform the analysis fixing $H_0=69.3\hu$ and $\alpha_M=0$.
When providing no redshift information of NGC4993, this reproduces\footnote{Note that for fixed $H_0$ our prior in redshift reproduce the $d^2$ prior used in \cite{2019PhRvL.123a1102A}.} exactly the analysis of \cite{2019PhRvL.123a1102A}.
When providing no redshift information we obtain $m^2_g=1.8_{-3.8}^{+11.7} \cdot 10^{-44} \rm{eV^2/c^4}$, while when we provide redshift information we obtain $m^2_g=1.7_{-3.1}^{+10.3} \cdot 10^{-44} \rm{eV^2/c^4}$. 
This not only shows that the improvement is due to the PN posterior samples themselves, but also that when providing the galaxy redshift information, one can tighten the bound on $m^2_g$ by 15\% when fixing $H_0$ and $\alpha_M$.

Below we provide results calculated from our inference on $\alpha_M$ in \textit{Scenario II} (for the massive gravity GW dispersion relation and running $H_0$).
The results that we obtain are compatible with previous results in \cite{2018JCAP...07..048P,2019PhRvL.123a1102A,2019PhRvD..99h3504L} even if we add a GW dispersion relation and can be converted into constraints on those theories which modify the GW luminosity distance as in Eq.~\eqref{eq:GW_lum}.  
For instance, for models with extra dimensions with the parametrisation $h\propto d_{\rm EM}^{-\gamma}$ with $\gamma=(D-2)/2$ \cite{2007ApJ...668L.143D,2018JCAP...07..048P,2019PhRvL.123a1102A}, we find the number of the spacetime dimensions to be $3.94<D<4.37$. 
For scalar-tensor theories with  running  Planck mass and no GW dispersion, $d_{\rm GW}/d_{\rm EM}=M_{\rm{Pl,eff}}({\rm today})/M_{\rm{Pl,eff}} (\rm {source})$ we obtain a value of $0.9<M_{\rm{Pl,eff}}({\rm today})/M_{\rm{Pl,eff}} (\rm {source})<1.8$, see also \cite{2019PhRvD..99h3504L}.
For non-local RR models \cite{2018JCAP...03..002B} with an effective Newton's constant we find $d^2_{\rm GW}/d^2_{\rm EM}=G ({\rm source})/G({\rm today})$ with $0.82<G ({\rm source})/G({\rm today})<3.34$. 
This can be converted into a constraint on $\dot{G} ({\rm source})/G({\rm today}$ using (\ref{eq:GW_lum}) and the Friedmann equation. We find $-2.5 \cdot 10^{-8} {\rm yr^{-1}}<\frac{G({\rm source})}{G({\rm today)}}<1.7 \cdot 10^{-9}{\rm yr^{-1}}$ at 95\% CL (here the upperbound is determined by $\alpha_M=-10$, the lowerbound of the prior on $\alpha_M$). This is competitive with other GW constraints, see e.g.~\cite{2020arXiv200312832V}, but is generally less stringent than solar system constraints \cite{2011LRR....14....2U}.
Complementary constraints on $\alpha_M$ are obtained from CMB, e.g.~\cite{2019arXiv190703150F,2019PhRvD..99j3502N}, assuming early time modifications of gravity.

\section{Conclusions \label{sec:6}}
In this paper we have presented a new method for probing deviations from GR at cosmological scales. 
In Sec.~\ref{sec:2}, we have discussed how in theories beyond GR, the GW propagation is modified in an expanding universe, and in particular we have shown how to relate the GW friction to a modified GW luminosity distance, and how to relate the GW speed (or dispersion relation) to the possible GW-GRB delay and GW phase evolution.  These 3 observables, which depend on the Hubble constant $H_0$, can contribute to measure GR deviations on cosmological scales. 

In Sec.~\ref{sec:3}, we discussed the required level of accuracy that we would need on the 3 observables to accurately measure GR deviations at a given redshift. 
Regarding the GW friction term $\alpha_M$, we have shown that the GW luminosity distance uncertainty is too high to allow for the accurate measurement of $\alpha_M$.
We have also discussed the possibility to constrain the GW dispersion parameter $\hat{\alpha}_j$ from the GW-GRB time delay and the GW phase independently.
We have shown that the GW phase  already provides one of the most stringent upperlimits on the GW dispersion relation.

In Sec.~\ref{sec:4} we have presented a Bayesian statistical framework able to combine GW, GRB and galaxy redshift measurements to jointly constrain the Hubble constant, the GW friction and the GW dispersion relation. 
The statistical method can be used starting from the posterior distributions of the redshift, GW parameters and GW-GRB time delay. 
Using this statistical framework, we have shown that our lack of knowledge on the value of the Hubble constant $H_0$ is a crucial variable to include if we are trying to measure the GW friction and dispersion relation. 
We have shown that by combining 100 BNSs events with accurate redshift estimations, $H_0$ will be constrained to an accuracy of 2 \% and the GW friction and dispersion relation (for the case of massive gravity) to an accuracy of 15\% and 2\% respectively. 
Note that these forecast are sensitive to the chosen type of GW dispersion relation given the accuracy on the corresponding PN order. 
For massive gravity, we have assumed that the 1PN coefficient could be constrained with $10\%$ accuracy. This resulted in a $H_0,\alpha_M$ and $\alpha_{-2}$ accuracy of 2\%,15\% and 2\%. However, higher PN coefficients are usually bounded with and accuracy 5-10 times worse\cite{2019PhRvL.123a1102A}. This means that, for those dispersion relations, the accuracy on the cosmological parameters will  be 5-10 times worse with 100 BNS events (if the posterior will reach Gaussian convergence). 
We argue that with the combination of 100 BNSs events, the error budget on $\alpha_M$ and $\hat{\alpha}_j$ is impacted by the $H_0$ determination and fixing an $H_0$ value would lead to a biased measurement of GR deviations.

Finally, we used the proposed framework to reanalyze GW170817 
adding $\aj$ over and above $(\alpha_M,H_0)$  and showing that this does not significantly modify the posteriors for either $H_0$ and $\alpha_M$.
We have shown that for GW170817 the error on $\hat{\alpha}_j$ is increased by $\sim 20\%$ if, rather than fixing $\alpha_M$, we marginalize over it. 
Using GW170817 we have also shown that for all the GW dispersion relations considered, the GW-GRB observation delay introduced by gravity modifications is negligible relative to the (redshifted) emission delay at the source. As a consequence, combining GW phase and timing of the GRB, one can accurately time the GRB emission at the source, concluding that the GRB is emitted exactly $1.72^{+0.05}_{-0.05}$s after the GW. 
This result can also be interpreted as a tighter constraint on the speed of GW at the merger frequency. Indeed, we find in the worst case (namely for $j=-1/3)$ $|v_g(f_{R,d})/c-1| \lesssim 10^{-17}$ while for massive gravity, where $j=-2$, $|v_g(f_{R,d})/c-1| \lesssim 5 \cdot 10^{-20}$. 
We stress that, though the GW phase is already setting a tight constraint on the GW dispersion relation, as remarked above, a timing of the GRB of the order of millisecond would improve the measurement of the dispersion relations corresponding to the PN  parameters 4, 5l, 6 and 7 as we show in Fig.~\ref{fig:tg_vg}.
\begin{acknowledgements}
We are grateful to G. Ashton, J.~M. Ezquiaga, A.~Ghosh,  N. Johnson-McDaniel, E. Maggio, M. Maggiore and A.~Samajdar for useful comments on this work.
We would like to acknowledge the support of the LabEx UnivEarthS (ANR-10-LABX-0023 and ANR-18-IDEX-0001), of the European Gravitational Observatory and of the Paris Center for Cosmological Physics.
DAS would like to thank the University of Geneva for hospitality whilst this work was in progress. 
\end{acknowledgements}

\begin{appendix}

\section{The selection function \label{app:A}}
The selection function $\beta$ appears at the denominator of the posterior distribution in Eq.~\ref{eq:pos}.
The $\beta$ function accounts for possible selection biases in the detection of events, i.e. for some choices of the parameters $(H_0,\alpha_M)$ events are more common to detect.
In principle, the selection function should be built from the knowledge of the detection probabilities $P_{\rm det}$ for the GW, GRB and hosting galaxy \cite{2019MNRAS.486.1086M}, i.e.
\begin{eqnarray}
    \beta(\alpha_M,H_0) &=& \int_0^{\infty} P_{\rm det}^{\rm GW}(\alpha_M,H_0,z)P_{\rm det}^{\rm GRB}(\alpha_M,H_0,z) \nonumber \\ && P_{\rm det}^{\rm galaxy}(\alpha_M,H_0,z) p(z|H_0) dz,
\end{eqnarray}
In practice, following \cite{2019PhRvD..99h3504L}, we compute $\beta(\alpha_M,H_0)$ as a function only of the detection probability of the GW event $P_{\rm det}^{\rm GW}$. 
This assumption is motivated by the fact that, for well localized BNS GW events, and any choice of the parameters $H_0,\alpha_M$, the GW maximum (also referred as ``horizon'') redshift for detection  $z_{H}^{\rm GW}$ is much lower than the GRB and hosting galaxy maximum redshift for detection $z_{H}^{\rm GRB/galaxy}$ \cite{2018Natur.562..545C,2019PhRvD..99h3504L}. 
With this approximation we can assume $P_{\rm det}^{\rm galaxy}(\alpha_M,H_0,z)=P_{\rm det}^{\rm GRB}(\alpha_M,H_0,z)=1$ for $z<z_{H}^{\rm GW} \ll z_{H}^{\rm GRB/galaxy}$. 
Therefore the selection function can be written as
\begin{equation}
\beta(\alpha_M,H_0) = \int_0^{z_H^{\rm GW}}  P_{\rm det}^{\rm GW}(\alpha_M,H_0,z) p(z|H_0) dz.
\label{eq:finsel}
\end{equation}

\begin{figure}[h!]
    \centering
    \includegraphics[scale=0.6]{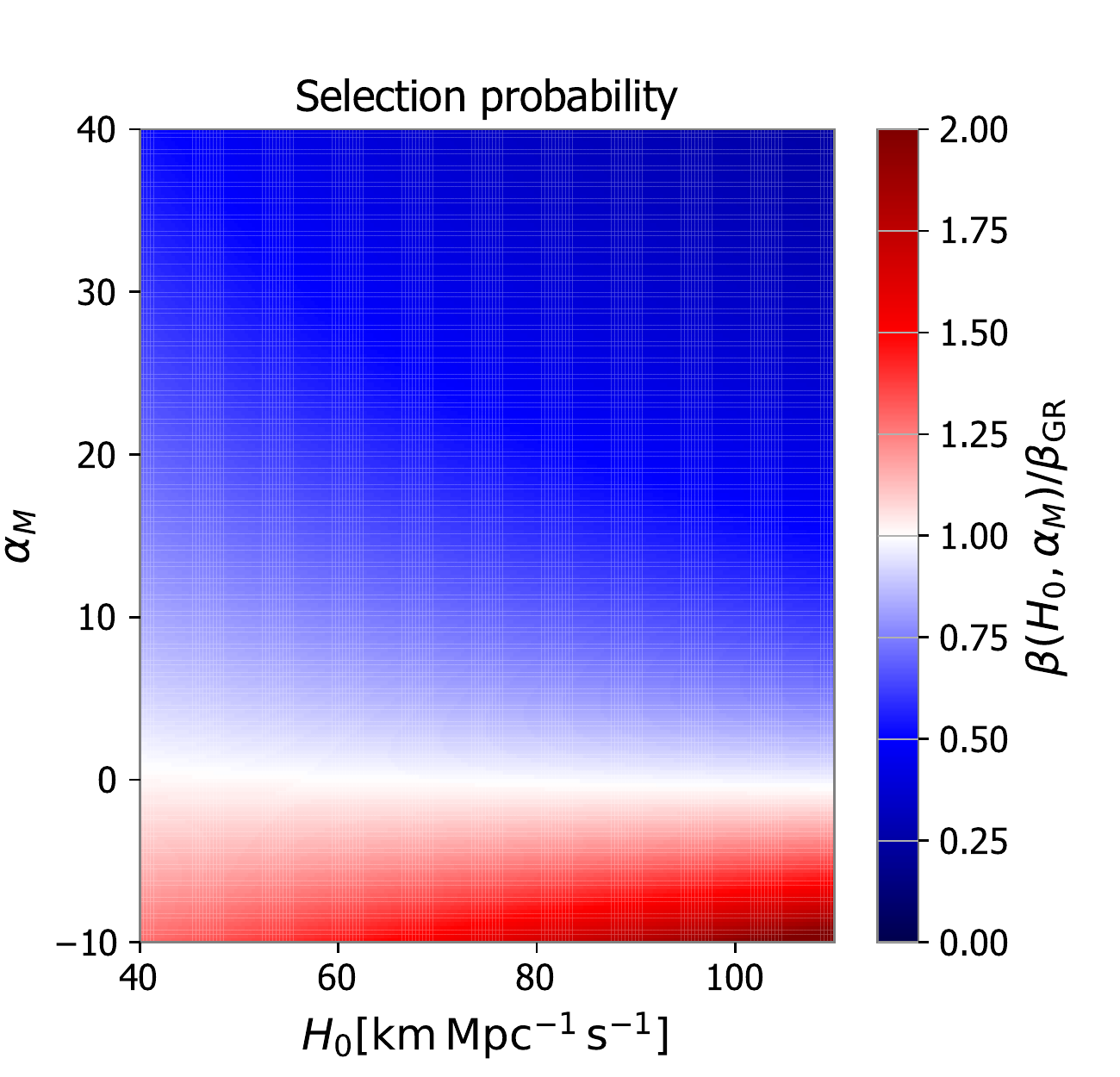}
    \caption{The colorbar displays the value of the selection function normalized over its value for $H_0=70 \hu$ and $\alpha_M=0$. Horizontal axis, Hubble constant. Vertical axis, GW friction parameter.}
    \label{fig:selection_probability}
\end{figure}

In order to compute $P_{\rm det}^{\rm GW}(\alpha_M,H_0,z)$ we use a Monte Carlo simulation.  
For each pair of $(\alpha_M,H_0)$ and varying $z$ from $0$ to $ z_{H}^{GW}$ in 1000 steps, we simulate 10000 BNS mergers with {\it (i)} an isotropic distribution for their sky-location {\it (ii)}  a uniform distribution of $\cos \iota$ (orbital inclination angle with respect to the line-of-sight) {\it (iii)} a chirp mass drawn from a gaussian distribution with mean $1.21 M_{\odot}$ and variance $0.20 M_{\odot}$.
Hence, we compute the optimal signal-to-noise ratio (SNR) for each GW event assuming LIGO Hanford, Livingston and Virgo detectors with power spectral densities (PSD) representative of O2 \footnote{https://www.gw-openscience.org/O2/}.
The GW detection probability is finally evaluated as the number of GW events recovered with an optimal SNR higher than $8$ over the total number of simulated binaries.
In Fig.~\ref{fig:selection_probability} we show the selection probability.
From the Figure we can see that the selection function is constant over $H_0$ for $\alpha_M=0$ (the General Relativity case).
This is because the integral in Eq.~\eqref{eq:finsel} is $\propto (z_{H}^{\rm GW}/H_0)^3$, and the GW horizon for detection $z_{H}^{\rm}$ scales as $H_0$ for low redshift. 
Hence, for $\alpha_M=0$ the selection is constant and there is no preference for the different values of $H_0$.
On the other hand if $\alpha_M \neq 0$, the GW redshift horizon will be a non-trivial function of $\alpha_M$ and $H_0$.
In general, for negative $\alpha_M$, the GWs will look closer and the GW horizon redshift for detection is increased, thus the selection function is higher.
On the other hand, for positive $\alpha_M$, GWs will look further and the GW horizon redsfhit for detection is reduced along with the selection function.
Eventually, positive values of $\alpha_M$ will be preferred by the selection function in Eq.~\eqref{eq:pos}.

\section{Monte Carlo Markov Chain setup \label{app:B}}
We use the Parallel Tempering Ensemble Monte Carlo Markov chain (PTEMCEE) of \cite{2013PASP..125..306F} implemented in the Python package {\it Bilby} \cite{2019ApJS..241...27A}.
As opposed to the classical Monte Carlo Markov chain codes, which use a single chain and require a proposal distribution for generating new posterior samples, the PTEMCEE iterates in parallel several independent chains which are used to auto-generate new posterior samples without the necessity of a proposal distribution.
We use 2 temperatures for the parallel tempering: a temperature of $1$ for the chain from which we sample and a warmer temperature of $10$ for the second chain. 
We set the parameter $a$ \cite{2013PASP..125..306F} for generating new posterior samples from the independent chains to its default value of $2$.
We run the PTEMCEE for 3000 iterations using 40 independent Markov chains.
At the end of the sampling, we discard the first 1000 iterations from all the chains.
This allow us to exclude posterior samples which are generated during the transient period of the PTEMCEE.
Therefore, we sub-sample the chains with a rate set equal to their auto-correlation time.

\end{appendix}

\bibliography{biblio}

\end{document}